\pdfoutput=1

\documentclass[11pt,reqno,preprint]{article}
\usepackage{jheppub}
\usepackage{epsfig}
\usepackage{amssymb}
\usepackage{amsmath}
\usepackage{mathrsfs}
\usepackage{hyperref}
\usepackage{multirow}
\usepackage{dsfont}
\usepackage{feynmp}
\usepackage{graphicx}
\usepackage[small,nohug,heads=vee]{diagrams}
\diagramstyle[labelstyle=\scriptstyle]

\def\be{\begin{equation}}
\def\ee{\end{equation}}
\def\ba{\begin{eqnarray}}
\def\ea{\end{eqnarray}}
\def\nl{\nonumber\\}

\def\CP1{\mathbb{CP}^1}
\def\SL2C{\mathrm{SL}(2,\mathbb{C})}

\def\Z2{\mathbb{Z}_2}

\title{Scattering of Massless Particles:\\Scalars, Gluons and Gravitons}
\author{Freddy Cachazo${}^{a}$, Song He${}^{a,b}$ and Ellis Ye Yuan${}^{a,c}$}
\affiliation[a]{Perimeter Institute for Theoretical Physics, Waterloo, ON N2L 2Y5,
Canada}
\affiliation[b]{School of Natural Sciences,
Institute for Advanced Study, Princeton, NJ 08540, USA}
\affiliation[c]{Department of Physics \& Astronomy, University of Waterloo, Waterloo, ON N2L 3G1,
Canada}
\emailAdd{fcachazo, she, yyuan@perimeterinstitute.ca}

\abstract{In a recent note we presented a compact formula for the complete tree-level S-matrix of pure Yang-Mills and gravity theories in arbitrary spacetime dimension. In this paper we show that a natural formulation also exists for a massless colored cubic scalar theory. In Yang-Mills, the formula is an integral over the space of $n$ marked points on a sphere and has as integrand two factors. The first factor is a combination of Parke-Taylor-like terms dressed with $U(N)$ color structures while the second is a Pfaffian. The S-matrix of a $U(N)\times U(\tilde N)$ cubic scalar theory is obtained by simply replacing the Pfaffian with a $U(\tilde N)$ version of the previous $U(N)$ factor. Given that gravity amplitudes are obtained by replacing the $U(N)$ factor in Yang-Mills by a second Pfaffian, we are led to a natural color-kinematics correspondence. An expansion of the integrand of the scalar theory leads to sums over trivalent graphs and are directly related to the KLT matrix. Combining this and the Yang-Mills formula we find a connection to the BCJ color-kinematics duality as well as a new proof of the BCJ doubling property that gives rise to gravity amplitudes. We end by considering a special kinematic point where the partial amplitude simply counts the number of color-ordered planar trivalent trees, which equals a Catalan number. The scattering equations simplify dramatically and are equivalent to a special Y-system with solutions related to roots of Chebyshev polynomials. The sum of the integrand over the solutions gives rise to a representation of Catalan numbers in terms of eigenvectors and eigenvalues of the adjacency matrix of an $A$-type Dynkin diagram.
}

\begin{document}
\maketitle

\section{Introduction and Summary of Results}

In 2003, Witten proposed a formula for the tree-level S-matrix of Yang-Mills in four dimensions, as an integral over the moduli space of certain rational maps from a $n$-punctured sphere to twistor space~\cite{Witten:2003nn}. Shortly after, Roiban, Spradlin and Volovich (RSV) studied the formula in momentum space and gave non-trivial evidence for its validity~\cite{Roiban:2004yf}. In 2012, an analogous construction for gravity in four dimensions was found~\cite{Cachazo:2012da,Cachazo:2012kg}. A natural question is whether similar constructions exist in arbitrary dimension.

In recent work~\cite{Cachazo:2013hca} we presented compact formulas for the complete tree-level S-matrix of Yang-Mills and gravity theories in any dimension. In their simplest form, both formulas can be written in a unified manner as
\be
{\cal M}^{({\bf s})}_n = \int\frac{d^n\sigma}{\textrm{vol}\,\SL2C}\prod_a {}'\delta(\sum_{b\neq a} \frac{s_{ab}}{\sigma_{a}-\sigma_{b}}) \left( \frac{{\rm Tr}(T^{\textsf{a}_1}T^{\textsf{a}_2}\cdots T^{\textsf{a}_n})}{(\sigma_{1}-\sigma_{2})\cdots (\sigma_{n}-\sigma_{1})}+\ldots \right)^{2-{\bf s}} \left({\rm Pf}'\Psi\right)^{\bf s}
\label{uni}\ee
with ${\bf s}=1$ for Yang-Mills and ${\bf s}=2$ for gravity. In this formula $T^{\textsf{a}}$ denotes the $U(N)$ color group generators, and the ellipsis means a sum over all permutations of labels modulo cyclic ones. Here $s_{ab}=(k_a+k_b)^2$ and by defining $e_{ab}=(\epsilon_a+\epsilon_b)^2$ and $d_{ab}=(\epsilon_a+k_b)^2$ with the understanding that the polarization vectors $\epsilon$'s are null, one can write $\Psi$, which is a $2n\times 2n$ antisymmetric matrix, as %that contains all the dependence on external momenta and polarization vectors,
\be \Psi_{a,b}=\begin{cases} \displaystyle \frac{s_{ab}}{\sigma_{a}-\sigma_{b}} & a\neq b,\\
\displaystyle \quad ~~ 0 & a=b,\end{cases} ~~ \Psi_{a{+}n, b{+}n} = \begin{cases} \displaystyle \frac{e_{ab}}{\sigma_{a}-\sigma_{b}} & a\neq b,\\
\displaystyle \quad ~~ 0 & a=b,\end{cases} ~~ \Psi_{a{+}n,b}=\begin{cases} \displaystyle \frac{d_{ab}}{\sigma_{a}-\sigma_{b}} & a\neq b,\\
\displaystyle -\sum_{c\neq a}\frac{d_{ac}}{\sigma_{a}-\sigma_{c}} &  a=b,\end{cases}
\ee
for $1\leq a,b\leq n$, and the block $\Psi_{a,b{+}n}$ follows from the antisymmetry of the matrix. ${\rm Pf}'\Psi\equiv \frac{(-1)^{i+j}}{\sigma_i{-}\sigma_j} {\rm Pf} \Psi^{i j}_{i j}$ for $1\leq i<j\leq n$ is called the reduced Pfaffian of $\Psi$, where $\Psi^{i j}_{i j}$ denotes the non-singular matrix obtained by removing columns $\{i,j\}$ and rows $\{i,j\}$ from $\Psi$~\footnote{Note that the formulas above differ from those in~\cite{Cachazo:2013hca} by some overall constant factors that can be absorbed into the definition of the coupling constants. More explicitly, ${\cal M}^{{\rm YM, here}}_n=\frac 1 2 {\cal M}^{{\rm YM, there}}_n$ and ${\cal M}^{{\rm gravity, here}}_n=2^{n{-}1} {\cal M}^{{\rm gravity, there}}_n$. The convention we use in this paper (which coincides with that in~\cite{Cachazo:2013gna}) is more standard, and we will see that it is convenient for connecting formulas with different $\textbf{s}$.}. The meaning of the symbol $\prod'$ and illustrations on how to use the formula explicitly are reviewed in section 2.

Here we would like to consider \eqref{uni} not only as a convenient way to write Yang-Mills and gravity scattering matrices in a unified way but also as a definition of the S-matrix for particles of spin ${\bf s}$. This means that the case ${\bf s}=0$ should correspond to a scalar theory. In this paper we show that this is indeed the case.

In order to make the claim more precise, recall that %in gravity the Pfaffian which appears squared can be regarded
the formula for pure gravity can be slightly generalized~\cite{Cachazo:2013hca} by replacing the integrand by the product of two independent Pfaffians, each with its own choice of gauge for polarization vectors
\be
({\rm Pf}'\Psi(\epsilon, k,\sigma ))^2 \rightarrow {\rm Pf}'\Psi(\epsilon, k,\sigma )\times {\rm Pf}'\Psi(\tilde\epsilon, k,\sigma ).
\ee
As proven in~\cite{Cachazo:2013gna}, the corresponding formula is obtained by applying the Kawai-Lewellen-Tye (KLT) relations to two copies of the Yang-Mills formula with polarizations $\epsilon$ and $\tilde\epsilon$. It is well known that the result gives amplitudes with gravitons coupled to dilatons and B-fields~\cite{Kawai:1985xq}.
%To avoid cluttering of equations let us denote ${\rm Pf}'\Psi(\epsilon, k,\sigma )$ by $E$ and ${\rm Pf}'\Psi(\tilde\epsilon, k,\sigma )$ by $\tilde E$.

This suggests that when we set ${\bf s}=0$ in \eqref{uni} to get a scalar theory, the integrand can also be generalized as
\be
\left( \frac{{\rm Tr}(T^{\textsf{a}_1}T^{\textsf{a}_2}\cdots T^{\textsf{a}_n})}{\sigma_{12}\sigma_{23}\cdots \sigma_{n1}}+\ldots \right)^2 \rightarrow \left( \frac{{\rm Tr}(T^{\textsf{a}_1}T^{\textsf{a}_2}\cdots T^{\textsf{a}_n})}{\sigma_{12}\sigma_{23}\ldots \sigma_{n1}}+\ldots \right) \left( \frac{{\rm Tr}({\tilde T}^{\textsf{b}_1}{\tilde T}^{\textsf{b}_2}\cdots {\tilde T}^{\textsf{b}_n})}{\sigma_{12}\sigma_{23}\cdots \sigma_{n1}}+\ldots \right)
\ee
where $\sigma_{ab}$ denotes $\sigma_a-\sigma_b$ (this notation is used here and in the rest of the paper in order to keep formulas more compact). Note that while the original squared factor depends on a single color group $U(N)$, the new factor has, in general, a different color group $U(\tilde N)$ with ${\tilde T}^{\textsf{b}}$ as its generators.

%Once again, in order to keep the notation simple we denote $C_{U(N)}$ the first factor and $C_{U(\tilde N)}$ the second.

This naturally motivates the study of a theory of scalars in the adjoint of the product of two different color groups $U(N)\times U(\tilde N)$. The simplest possibility is the theory with only cubic interactions of the form,
\be
-f_{\textsf{abc}}\tilde f_{\textsf{a'b'c'}}\phi^{\textsf{a}\textsf{a'}}\phi^{\textsf{b}\textsf{b'}}\phi^{\textsf{c}\textsf{c'}}
\ee
where $f_{\textsf{abc}}$ and $\tilde f_{\textsf{a'b'c'}}$ are structure constants of $U(N)$ and $U(\tilde N)$ respectively.

In this note we show that ${\cal M}^{(0)}_n$ defined as
\be
\int\!\!\frac{d^n\sigma}{\textrm{vol}\,\SL2C}\prod_a {}'\delta(\sum_{b\neq a} \frac{s_{ab}}{\sigma_{a}-\sigma_{b}}) \left( \frac{{\rm Tr}(T^{\textsf{a}_1}T^{\textsf{a}_2}\cdots T^{\textsf{a}_n})}{\sigma_{12}\sigma_{23}\ldots \sigma_{n1}}+\ldots \right) \left( \frac{{\rm Tr}({\tilde T}^{\textsf{b}_1}{\tilde T}^{\textsf{b}_2}\cdots {\tilde T}^{\textsf{b}_n})}{\sigma_{12}\sigma_{23}\cdots \sigma_{n1}}+\ldots \right)
\label{colorscalar}
\ee
gives the full tree-level S-matrix of such scalar theory in any dimension.

%We define the S-matrix as the sum of Feynman diagrams with an overall minus sign.

All of the above also leads to the natural conclusion that the factors
\be
C_{U(N)} \equiv \left( \frac{{\rm Tr}(T^{\textsf{a}_1}T^{\textsf{a}_2}\cdots T^{\textsf{a}_n})}{\sigma_{12}\sigma_{23}\ldots \sigma_{n1}}+\ldots \right) \quad {\rm and} \quad E_{\epsilon} \equiv {\rm Pf}'\Psi (\epsilon)~\label{colorkinematicfactors}
\ee
are interchangeable as they give rise to physical theories in the process. This is a color-kinematics correspondence which is
valid for individual solutions to the scattering equations.

More precisely, the formulas with ${\bf s}=0,1,2$ are closely related.
%Let us denote the combinations inside the two brackets in eq.~\eqref{colorscalar} as $C$ and $\tilde C$, and the reduced Pfaffians in the gravity formula as $E$ and $\tilde E$.
Starting from the formula for the scalar theory, with integrand $C_{U(N)}\times C_{U(\tilde N)}$, if we replace  $C_{U(N)}$ (or $C_{U(\tilde N)}$), by the Pfaffian $E_\epsilon$ (or $E_{\tilde\epsilon}$), we get the Yang-Mills formula with color group $U(\tilde N)$ (or $U(N)$); if we further replace $C_{U(\tilde N)}$ (or $C_{U(N)}$) in the Yang-Mills formula by another copy of the Pfaffian $E_{\tilde\epsilon}$ (or $E_\epsilon$), we arrive at the gravity formula. We summarize the relations by the following diagram
%{\cal M}^{(0)}_n (U(N)\times U(\tilde N)) \,\xrightarrow{\smash{C\to E}\, (\tilde C\to \tilde E)}\,{\cal M}^{(1)}_n ( U(\tilde N))\,({\cal M}^{(1)}_n ( U(N))) \,\xrightarrow{\smash{\tilde C\to \tilde E}\,(C \to E)}\,{\cal M}^{(2)}_n.

\begin{displaymath}
\begin{diagram}
%&&&&{\cal M}^{(1)}_n ( U(N), \tilde E)\\
%&&&\ruTo^{\tilde C\to \tilde E} &&\rdTo^{C\to E}\\
%&{\cal M}^{(0)}_n (U(N)\times U(\tilde N)) &&&&&&{\cal M}^{(2)}_n (E,\tilde E)\\
%&&&\rdTo_{C \to E} &&\ruTo_{\tilde C\to \tilde E}\\
%&&&&{\cal M}^{(1)}_n ( U(\tilde N), E)
&&&&{\cal M}^{(1)}_n ( U(N), \tilde\epsilon)&\\
&&\ruTo^{C_{U(\tilde N)}\to E_{\tilde\epsilon}} &&&&\rdTo^{C_{U(N)}\to E_{\epsilon}}\\
{\cal M}^{(0)}_n (U(N)\times U(\tilde N)) &&&&&&&&{\cal M}^{(2)}_n (\epsilon,\tilde \epsilon)&\\
&&\rdTo_{C_{U(N)}\to E_{\epsilon}} &&&&\ruTo_{C_{U(\tilde N)}\to E_{\tilde\epsilon}}\\
&&&&{\cal M}^{(1)}_n ( U(\tilde N), \epsilon)&\\
\end{diagram}
\end{displaymath}

Amplitudes in theories with color can be decomposed into partial amplitudes, dressed with color factors.
For the scalar theory with color group $U(N)\times U(\tilde N)$, one can decompose ${\cal M}^{(0)}_n$ with respect to either copy of the color groups, e.g. $U(\tilde N)$, in terms of the traces in~\eqref{colorscalar}
\be\label{chaindecomposition1}
{\cal M}^{(0)}_n =\sum_{\alpha\in S_{n}/Z_{n}} {\rm Tr}({\tilde T}^{\textsf{b}_{\alpha(1)}}{\tilde T}^{\textsf{b}_{\alpha(2)}}\cdots {\tilde T}^{\textsf{b}_{\alpha(n)}}) M^{(0)}_n(\alpha(1),\alpha(2),\ldots,\alpha(n)),
\ee
where $M^{(0)}_n(\alpha(1),\alpha(2),\ldots,\alpha(n))\equiv M^{(0)}_n(\alpha)$ are known in the literature as color-ordered partial amplitudes. Furthermore, one can decompose $M^{(0)}_n(\alpha)$ with respect to the other copy of color group, $U(N)$,
\be\label{chaindecomposition2}
M^{(0)}_n(\alpha)=\sum_{\beta\in S_{n}/Z_{n}} {\rm Tr}(T^{\textsf{a}_{\beta(1)}}T^{\textsf{a}_{\beta(2)}}\cdots T^{\textsf{a}_{\beta(n)}}) m^{(0)}_n (\alpha|\beta),
\ee
into what we call `double-partial' amplitudes, denoted as $m^{(0)}_n(\alpha|\beta)$.

There are two natural specializations of our formula which give rise to the two cases of partial amplitudes. The first case is (for simplicity we write the explicit formula with the canonical ordering $\alpha=I$), %these two interesting sums over Feynman diagrams.
\be
M^{(0)}_n=\! \int\!\!\frac{d^n\sigma}{\textrm{vol}\,\SL2C}\prod_a {}'\delta(\sum_{b\neq a} \frac{s_{ab}}{\sigma_{ab}})\left(\sum_{\beta\in S_{n}/Z_{n}}\frac{{\rm Tr}(T^{\textsf{a}_{\beta(1)}}T^{\textsf{a}_{\beta(2)}}\cdots T^{\textsf{a}_{\beta(n)}})}{\sigma_{\beta(1),\beta(2)}\cdots\sigma_{\beta(n),\beta(1)}}\right)\frac{1}{\sigma_{1,2}\cdots\sigma_{n,1}}.
\label{colored}\ee
By definition $M^{(0)}_n$ must be given by the sum over trivalent $U(\tilde N)$-color-ordered Feynman diagrams where each vertex is dressed with a structure constant $f_{\textsf{abc}}$ of $U(N)$.

The second one gives double-partial amplitudes
\be
m^{(0)}_n(\alpha|\beta)\!=\!\!\int \frac{d\,^n\sigma}{\textrm{vol}\,\SL2C}\frac{\prod_a {}'\delta(\sum_{b\neq a} \frac{s_{ab}}{\sigma_{a b}})}{(\sigma_{\alpha(1),\alpha(2)}\cdots\sigma_{\alpha(n),\alpha(1)})
(\sigma_{\beta(1),\beta(2)}\cdots\sigma_{\beta(n),\beta(1)})}
\label{scal}.
\ee
%
%\ba
%m^{(0)}_n(\alpha|\beta)&=& \int \frac{d\,^n\sigma}{\textrm{vol}\,\SL2C}\prod_a {}'\delta(\sum_{b\neq a} \frac{k_a\cdot k_b}{\sigma_{a}-\sigma_{b}})\nl
%&&\times \frac{1}{(\sigma_1{-}\sigma_{\alpha(2)})\ldots(\sigma_{\alpha(n{-}1)}{-}\sigma_n)(\sigma_n{-}\sigma_1)(\sigma_1{-}\sigma_{\ \beta(2)})\ldots(\sigma_{\beta(n{-}1)}{-}\sigma_n)(\sigma_n{-}\sigma_1)}.
%\label{scal}.
%\ea
In section 3, we show that $m^{(0)}_n(\alpha|\beta)$ is given by the sum over all trivalent graphs which have two planar embeddings, one consistent with the $\alpha$ ordering and the other with the $\beta$ ordering. In the special case $\alpha=\beta$, the formula gives rise to the sum over all planar (ordered) trivalent graphs with weights given by the product of all scalar propagators. %The double-partial amplitudes $m^{(0)}_n(\alpha|\beta)$ have a direct connection to the KLT bilinear, which will also be discussed in Section 3.
In section 5 we focus on the kinematic regime where each trivalent graph evaluates to unity. This means the double-partial amplitude with $n$ particles evaluates to the total number of planar trivalent graphs which is $C_{n-2}$, the $(n-2)^{\rm th}$ Catalan number. We find that on this special kinematics the scattering equations simplify dramatically and are reduced to a special Y-system whose solutions are related to the roots of Chebyshev polynomials.%of the second kind.
We make a conjecture for what the integrand of our formula \eqref{scal} gives when evaluated on each solution.

Having a direct connection to scalar trivalent graphs and the fact that the color and kinematic factors in \eqref{colorkinematicfactors} are interchangeable indicates a link to the color-kinematics duality discovered by Bern, Carrasco and Johansson (BCJ) in 2008 \cite{Bern:2008qj}. In fact, we are able to show that an expansion of the kinematic factor $E_{\epsilon}$ analogous to that of the color factor $C_{U(N)}$ exists. This leads to a formula for the Yang-Mills amplitude as a linear combination of double-partial amplitudes, $m^{(0)}_n (\alpha|\beta)$. Using their expansion in terms of trivalent graphs, our formula leads directly to the BCJ color-kinematics duality. Furthermore, by using our transformation from Yang-Mills amplitudes to gravity amplitudes one finds a new proof for the BCJ double-copy relations~\cite{Bern:2008qj}. These facts will be discussed in Section 4.

Section 6 is devoted to consistency checks of our formula by showing that \eqref{colorscalar} has correct soft limits and factorization limits. There we also give a proof for the structure of the double-partial amplitudes \eqref{scal}. We end in section 7 with conclusions and discussions.

\section{Details of the Formula and Examples}

Let us discuss the precise definition of all the elements entering \eqref{scal} and then show how explicit computations are carried out. Let us denote by $\{ \sigma_1,\sigma_2,\ldots, \sigma_n\}$ the position of $n$ punctures in the complex plane. Below we will see that our formula has an $\SL2C$ invariance which means that we are dealing with $n$ punctures on $\mathbb{CP}^1$. One of the main ingredients of the formula are the scattering equations~\cite{Cachazo:2013iaa,Cachazo:2013gna}
\be
\sum_{b\neq a} \frac{s_{ab}}{\sigma_{a}-\sigma_{b}} = 0 \quad {\rm for} \quad a\in \{ 1,2,\ldots , n\}
\ee
which connect the space of kinematic invariants defined by scalar products of the momenta of external particles $\{k^\mu_a\}$ and the puncture locations $\{ \sigma_a\}$. Using the fact that in any physical process momentum is conserved and all particles are on-shell (i.e. $k^2_a=0$), one can show that the equations are $\SL2C$ invariant and only $n-3$ are linearly independent. Here $\SL2C$ acts as usual
\be
\sigma \to \frac{\textsc{a} \sigma+\textsc{b}}{\textsc{c}\sigma +\textsc{d}} \quad {\rm with} \quad \textsc{a}\textsc{d}-\textsc{b}\textsc{c} = 1.
\ee

A simple way of imposing the support of the scattering equations is by noticing that
\be
\prod_a {}'\delta(\sum_{b\neq a} \frac{s_{ab}}{\sigma_{a b}}) \equiv \sigma_{ij}\sigma_{jk}\sigma_{ki}\prod_{a\neq i,j,k}\delta(\sum_{b\neq a} \frac{s_{ab}}{\sigma_{a b}})
\ee
is independent of the choice $\{i,j,k\}$ and hence permutation invariant~\cite{Cachazo:2013gna}. %Here and in the rest of this paper $\sigma_{ab} = \sigma_a-\sigma_b$.

In Section 1, we discussed color decomposition by using traces of products of color group generators. However, in practice it is more convenient to use an alternative color basis proposed in~\cite{DelDuca:1999rs} which contains only $(n-2)!$ elements. It fixes the position of two particles, e.g. $1$ and $n$ while permuting the remaining $n-2$ labels. The traces are then replaced by
\be
{\bf c}_{\alpha}\equiv \sum_{\textsf{c}_1,\ldots,\textsf{c}_{n{-}3}} f_{\textsf{a}_1 \textsf{a}_{\alpha(2)} \textsf{c}_1}\cdots f_{\textsf{c}_{n{-}3} \textsf{a}_{\alpha(n{-}1)} \textsf{a}_n}, \label{c-basis}
\ee
where $\alpha\in S_{n{-}2}$, and similarly for $\tilde {\bf c}_{\alpha}$. $C_{U(N)}$ and $C_{U(\tilde N)}$ can be decomposed in terms of ${\bf c}_{\alpha}$ and $\tilde {\bf c}_{\alpha}$ respectively. In this color basis, partial amplitudes $M^{(0)}_n(\alpha)$ and double-partial amplitudes $m^{(0)}_n(\alpha|\beta)$ are identical to those in the trace basis, but with the position of $1$ and $n$ fixed.

Combining these with the Faddeev-Popov Jacobian obtained by fixing the $\SL2C$ redundancy acting on the $\sigma's$ one finds that the scalar amplitude, \eqref{colorscalar}, becomes
\be
%{\cal M}^{(0)}_n= \sum_{\{\sigma\} \in {\rm solutions}}\frac{1}{{\rm det}'\Phi}\left( \frac{{\rm Tr}(T^{\textsf{a}_1}T^{\textsf{a}_2}\cdots T^{\textsf{a}_n})}{(\sigma_{1}-\sigma_{2})\ldots (\sigma_{n}-\sigma_{1})}+\ldots \right) \left( \frac{{\rm Tr}({\tilde T}^{\textsf{b}_1}{\tilde T}^{\textsf{b}_2}\cdots {\tilde T}^{\textsf{b}_n})}{(\sigma_{1}-\sigma_{2})\ldots (\sigma_{n}-\sigma_{1})}+\ldots \right)
{\cal M}^{(0)}_n= \sum_{\{\sigma\} \in {\rm solutions}}\frac{1}{{\rm det}'\Phi}\sum_{\alpha,\beta\in S_{n{-}2}}\frac{{\bf c}_\alpha \tilde {\bf c}_\beta}{(\sigma_{\alpha(1),\alpha(2)}\cdots \sigma_{\alpha(n),\alpha(1)})(\sigma_{\beta(1),\beta(2)}\cdots\sigma_{\beta(n),\beta(1)})}\label{exp}\ee
where in the new color basis $\alpha(1)=\beta(1)=1,\alpha(n)=\beta(n)=n$, and the sum is over all the solutions to the scattering equations and
\be
\Phi_{ab} = \begin{cases}
\displaystyle \quad\frac{s_{ab}}{(\sigma_a-\sigma_b)^2} & a\neq b,\\
\displaystyle -\sum_{c\neq a}\frac{s_{ac}}{(\sigma_a-\sigma_c)^2} & a=b.
\end{cases}
\ee
is a corank $3$ matrix. To get the reduced determinant ${\rm det}'\Phi$, one removes any three rows $\{ i,j,k\}$ and any three columns $\{ p,q,r\}$ to get a reduced matrix, whose determinant, which we denote as $|\Phi|^{ijk}_{pqr}$, is non-vanishing. Then
\be
{\det}'\Phi \equiv  \frac{|\Phi|^{ijk}_{pqr}}{(\sigma_{pq}\sigma_{qr}\sigma_{rp})(\sigma_{ij}\sigma_{jk}\sigma_{ki})}.
\label{perf2}\ee
This matrix was first encountered in \cite{Cachazo:2012da} as a natural analog of Hodge's MHV gravity formula \cite{Hodges:2012ym}.

Finally, it is important to mention that an inductive algorithm for solving the scattering equations is known \cite{Cachazo:2013gna}. The total number of solutions for generic kinematics is $(n-3)!$. In section 5 we show that in some special situations the equations simplify dramatically and all solutions can be obtained analytically.

%As discussed in~\cite{Cachazo:2013gna}, formulas for Yang-Mills and gravity can be written in a similar fashion as summing over $(n-3)!$ solutions; in the summand, the reduced Pfaffian can also be expanded as a combination of Parke-Taylor like factors with coefficients being kinematic factors (i.e. functions of $k$ and $\epsilon$). This is similar to $C$ and $\tilde C$ which are combinations of Parke-Taylor like factors with color factors as coefficients, which can be viewed as an analog of color-kinematic duality~\cite{Bern:2008qj} in our formulation.

%Our observation above applies to each solution: for the summand in \eqref{exp} evaluated at the $S^{\text{th}}$ solution ($S \in \{1,2,\ldots,(n-3)!\}$), by replacing the color combination $C_S$ (or $\tilde C_S$), by the kinematic combination $E_S$ ($\tilde E_S$), it gives the summand for Yang-Mills amplitude, and by replacing both color combinations $C_S, \tilde C_S$ by $E_S,\tilde E_S$, it gives the summand for gravity amplitude. It is tempting to interpret this fact as an analog of the double-copy relations between Yang-Mills and gravity amplitudes~\cite{Bern:2008qj}, as well as an extension to the scalar theory. We leave further studies of the relations to future works.

\subsection{Examples}\label{sec:examples}

The simplest example is the three particle amplitude. In this case one finds that no equations have to be solved as all $\sigma$ variables can be fixed using the $\SL2C$ invariance. Moreover, ${\det}'\Phi = 1/ (\sigma_{12}\sigma_{23}\sigma_{31})^2$ which cancels the Parke-Taylor like factor squared in the integrand of \eqref{exp} to give
\be
{\cal M}^{(0)}_3 (1^\textsf{aa'},2^\textsf{bb'},3^\textsf{cc'})= f_{\textsf{abc}}\tilde f_{\textsf{a'b'c'}}.
\ee
which is the correct answer for a cubic scalar theory.

Next we compute the four particle amplitude. The scattering equations become a single equation for one variable. Solving the scattering equations with $\sigma_1=0$, $\sigma_2=1$, $\sigma_3=\infty$ gives $\sigma_4 = s_{24}/s_{34}$. Let us define $s_{12}=s, s_{23}=t$, $s_{13}=u$, and for four particles, we adopt the standard convention for labelling the color factors using the $s,t,u$ channels:

\be
{\bf c}_s=\sum_\textsf{b} f_{\textsf{a}_1\textsf{a}_2 \textsf{b}}f_{\textsf{b}\textsf{a}_3\textsf{a}_4},
\quad{\bf c}_t=\sum_\textsf{b} f_{\textsf{a}_1\textsf{a}_4\textsf{b}} f_{\textsf{b}\textsf{a}_3\textsf{a}_2},
\quad{\bf c}_u=\sum_\textsf{b} f_{\textsf{a}_1\textsf{a}_3\textsf{b}}f_{\textsf{b}\textsf{a}_2\textsf{a}_4}, \ee
and similarly for $\tilde{\bf c}_s,\tilde{\bf c}_t, \tilde{\bf c}_u$. We also denote the ordering $(1324)$ as $P$ (the canonical ordering is denoted as $I$).
Computing ${\det}'\Phi= |\Phi|_{234}^{123}/(\sigma_{12}\sigma^2_{23}\sigma_{31}\sigma_{34}\sigma_{42})$ and plugging the solution into the amplitude one gets
%
%\be
%{\cal M}^{(0)}_4 = \frac{\sigma_{31}\sigma_{42}}{\sigma_{12}\sigma_{34}}\frac{1}{s_{14}}.
%\ee
%
\ba
{\cal M}^{(0)}_4 %(1^\textsf{aa'},2^\textsf{bb'},3^\textsf{cc'},4^\textsf{dd'})
&=& {\bf c}_s \tilde{\bf c}_s m^{(0)}_4(I;I)+{\bf c}_s\tilde{\bf c}_u m^{(0)}_4 (I;P)+{\bf c}_u \tilde{\bf c}_s m^{(0)}_4(P;I)+{\bf c}_u \tilde{\bf c}_u m^{(0)}_4 (P;P)\nl
&=& {\bf c}_s\tilde{\bf c}_s \frac{u}{st} +({\bf c}_s \tilde{\bf c}_u +{\bf c}_u \tilde{\bf c}_s) \frac {1}{t}+{\bf c}_u \tilde{\bf c}_u \frac{s}{u t}\nl
&=& -\frac{{\bf c}_s \tilde{\bf c}_s}{s}-\frac{{\bf c}_t\tilde{\bf c}_t}{t}-\frac{{\bf c}_u\tilde{\bf c}_u} {u}
\ea
as expected for a color-dressed cubic theory amplitude. In the last equality we have used the Jacobi identities ${\bf c}_s -{\bf c}_u -{\bf c}_t=\tilde{\bf c}_s-\tilde{\bf c}_u -\tilde{\bf c}_t=0$.

We have also checked explicitly that formula \eqref{colorscalar} gives the correct five point amplitude. The full amplitude can be written as a term proportional to the partial amplitude $M^{(0)}_5(1,2,3,4,5)$, plus five terms related by permutations, and each partial amplitude can be decomposed into six double-partial amplitudes,
\be {\cal M}^{(0)}_5=\tilde {\bf c}_I \left(\sum_{i=0}^5 {\bf c}_{P_i} m^{(0)}_5(I|P_i)\right)+\textrm{permutations of}\,(2,3,4),\label{5ptfull}\ee
where we have denoted the orderings as $I=P_0$, $(13245)=P_1$, $(12435)=P_2$, $(14325)=P_3$, $(13425)=P_4$, $(14235)=P_5$. Let us first consider how to compute these double-partial amplitudes. In arbitrary dimensions the scattering equations give rise to an irreducible quadratic polynomial. We have checked that by summing over solutions as dictated by \eqref{exp} one reproduces the desired answer for the pair of two canonical orderings,
\be
m^{(0)}_5 (I|I)= \frac{1}{s_{12}s_{34}}+\frac{1}{s_{23}s_{45}}+\frac{1}{s_{34}s_{51}}+\frac{1}{s_{45}s_{12}}+\frac{1}{s_{51}s_{23}},
\label{5ptcanonical}
\ee
and for other five pairs of orderings needed in \eqref{5ptfull},
\be  m^{(0)}_5 (I|P_1)=-\frac 1 {s_{23}} (\frac1 {s_{45}}+\frac 1 {s_{51}}), ~ m^{(0)}_5 (I|P_2)=-\frac 1 {s_{34}} (\frac 1 {s_{51}}+\frac 1 {s_{12}}),~ m^{(0)}_5 (I|P_3)=-\frac 1 {s_{51}}(\frac 1 {s_{23}}+\frac 1 {s_{34}}),\nonumber\ee
\be m^{(0)}_5 (I|P_4)=-\frac 1 {s_{34}s_{51}}, \quad m^{(0)}_5 (I|P_5)=-\frac 1 {s_{23}s_{51}}. \label{5ptothers}\ee
By plugging them into the combination inside the bracket of \eqref{5ptfull} and after using Jacobi identities repeatedly, one gets the expected partial amplitude $M^{(0)}_5(1,2,3,4,5)$. By summing over permutations, \eqref{5ptfull} gives the full five-point amplitude.

In four dimensions, something special happens and when using spinor variables the quadratic polynomial from the scattering equations factorizes. In the terminology of Yang-Mills theory, one solution gives rise to the MHV sector while the other gives the $\overline{\textrm{MHV}}$ sector. In our computation we have to add up both solutions. It is easy to find the explicit formula for an arbitrary choice of permutations, $m^{(0)}_5(\alpha|\beta)$ to get
\be
\!\!\!\left(\frac{\prod_{i<j}\langle i~j\rangle}{\langle \alpha(1)\alpha(2)\ldots \alpha(5)\rangle\langle \beta(1)\beta(2)\ldots \beta(5)\rangle} -  \frac{\prod_{i<j}[ i~j ]}{[ \alpha(1)\alpha(2)\ldots \alpha(5)][ \beta(1)\beta(2)\ldots \beta(5)]}\right)\!\frac{1}{\epsilon(1234)}
\label{der}\ee
with
$$\langle 12345 \rangle=\langle 12\rangle\langle 23\rangle\cdots \langle 51\rangle,~~ [12345]=[12][23]\cdots [51]$$
and $\epsilon(1234) = \langle 12\rangle [23] \langle 34\rangle [41] - [12] \langle 23\rangle [34] \langle 41\rangle$.

This formula indeed reproduces \eqref{5ptcanonical} and \eqref{5ptothers}.

Finally, we have checked that our formula also reproduces the sum of planar trivalent diagrams up to eight particles. Just as in previous cases, it is crucial to sum over all $(n-3)!$ solutions to find the amplitude. 

In the next section we give an interpretation of each $m^{(0)}_n(\alpha|\beta)$ as a sum over trivalent graphs. Hints of what the interpretation is can already be obtained from the examples given in this section. One more example worth mentioning is the case $m^{(0)}_5(12345|13524)$. The reason this is special is that the two permutations shown as a graph connecting five points do not share any edges and therefore their union gives the complete graph with five vertices. Moreover, this means that, e.g., in four dimensions,
\be
\langle 12345\rangle\langle 13524\rangle = -\prod_{i<j}\langle i~j\rangle\quad {\rm and} \quad [12345][13524] = -\prod_{i<j}[i~j].
\ee
Plugging this into \eqref{der} immediately gives $m^{(0)}_5(12345|13524) = 0$.

\section{Double-Partial Amplitudes}\label{generatingfunctions}

Double-partial amplitudes are the building blocks of scalar amplitudes. In section 4, we show that they are also building blocks of Yang-Mills and gravity amplitudes. In this section, we show that double-partial amplitudes can be expanded as sums of trivalent graphs. Also, using a property of scattering equations called KLT orthogonality, we prove that the matrix with double-partial amplitudes as entries equals the inverse of the KLT matrix. 

\subsection{Trivalent Graph Expansion}

Here we study double-partial amplitudes as generating functions of sums of scalar diagrams. We first recall the formula defining double-partial amplitudes
\be\label{objectsdef}
m^{(0)}_n(\alpha|\beta)=\sum_{\{\sigma\}\in\text{solutions}}\frac{1}{(\sigma_{\alpha(1),\alpha(2)}\cdots\sigma_{\alpha(n),\alpha(1)})(\sigma_{\beta(1),\beta(2)}\cdots\sigma_{\beta(n),\beta(1)}){\det}'\Phi}.
\ee
The sum is over all $(n-3)!$ solutions to the scattering equations.

The simplest examples are those for $n=3,4$ and $n=5$ given in the previous section. Let us recall some of the results in order to motivate the proposal for the meaning of $m^{(0)}_n(\alpha|\beta)$. Consider
\be\label{simpleexamplesD}
m^{(0)}_3(I|I)=1,\quad
m^{(0)}_4(I|I)=-\frac{1}{s_{1 2}}-\frac{1}{s_{1 4}},\quad
m^{(0)}_4(I|1,3,2,4)=\frac{1}{s_{1 4}},
\ee
and
\be
m^{(0)}_5 (I|I)= \frac{1}{s_{12}s_{34}}+\frac{1}{s_{23}s_{45}}+\frac{1}{s_{34}s_{51}}+\frac{1}{s_{45}s_{12}}+\frac{1}{s_{51}s_{23}},
\nonumber\ee
\be
m^{(0)}_5 (I|13245)=-\frac 1 {s_{23}s_{45}}-\frac 1 {s_{23}s_{51}}, \quad m^{(0)}_5(I|13524) = 0.
\ee
From these examples it is easy to see that when both permutations in $m^{(0)}_n(\alpha|\beta)$ are the same then the answer is a sum over all color-ordered trivalent graphs, each contributing the product of its propagators. When the two permutations are different it gives a subset of terms appearing in the formula for $m(\alpha|\alpha)$. In extreme cases like $m^{(0)}_5(I|13524)$ the subset is the empty set. A closer look at the relation between the diagrams that contribute to a given case straightforwardly motivates the following proposition.

\vskip0.07in

{\bf Proposition:} {\it %The function $m^{(0)}_n(\alpha|\alpha)$ computes the sum over all $\alpha$-color-ordered trivalent graphs, each contributing the product of all its propagators. In general,
The function $m^{(0)}_n(\alpha|\beta)$ computes the sum of the collection of all trivalent scalar diagrams that can be regarded both as $\alpha$-color-ordered and $\beta$-color-ordered, where each diagram's contribution is given by the product of its propagators. In other words, only diagrams that belong to the intersection of both sets contribute to the double-partial amplitude.}

\vskip0.07in

More explicitly, let $\mathcal{T}(\alpha)$ denote the set of $\alpha$-color-ordered diagrams and $\mathcal{T}(\beta)$ the set of $\beta$-color-ordered ones. Then
\be\label{sumscalardiagrams}
m^{(0)}_n(\alpha|\beta)=(-1)^{n-3+n_{\text{flip}}(\alpha|\beta)}\!\!\sum_{g\in \mathcal{T}(\alpha)\cap \mathcal{T}(\beta)}~~\prod_{e\in E(g)}\frac{1}{s_e},
\ee
where the integer $n_{\text{flip}}(\alpha|\beta)$ is defined below and $s_e=P_e^2$ where $P_e$ is the momentum flowing along the edge $e$ in the set of edges, $E(g)$, for the Feynman diagram $g$. In particular, whenever $\mathcal{T}(\alpha )\cap \mathcal{T}(\beta )=\varnothing$ then $m^{(0)}_n(\alpha|\beta)=0$.

We give a proof of this proposition in section 6 by showing that $m^{(0)}_n(\alpha|\beta)$ has the same soft limits and factorizations properties as the sum over the corresponding Feynman diagrams. In the rest of this section we present an efficient description of the diagrams that appear in a particular amplitude as well as a procedure to determine $n_{\text{flip}}(\alpha|\beta)$.

Consider a particular double-partial amplitude $m^{(0)}_n(\alpha|\beta)$, without loss of generality take $\alpha$ to be the identity permutation, i.e., $\alpha = I$. Start by drawing a disk with $n$ nodes sitting on the boundary in the ordering $\alpha$. Then link the $n$ nodes together with a loop of line segments according to the ordering $\beta$. Generally the segments intersect each other in the middle of the graph. It is convenient to introduce the following terminology: A subset of all $n$ points which are consecutive with respect to the $\alpha$  (or $\beta$) ordering will be said to be $\alpha$-consecutive (or $\beta$-consecutive).

The way to compute $m^{(0)}_n(\alpha|\beta)$ is iteratively: Start by locating a set of at least two external labels which are both $\alpha$- and $\beta$-consecutive, say $\{i,i+1,\ldots, i+r\}$ with $r>1$ (if no set can be found then $m^{(0)}_n(\alpha|\beta)$ vanishes and if $r=n$ then $\beta = \alpha$ and $m^{(0)}_n(\alpha|\alpha)$ is given as above). If the set can be extended, e.g., by adding $i-1$ while still remaining $\beta$-consecutive then take its maximal extension.

Assuming $\{i,i+1,\ldots, i+r\}$ is maximal already, redraw the graph by moving all points in the set along the boundary of the disk, until they are close to each other\footnote{In order to make this precise one has to take the limit in which they become a single point. However, the more informal description is enough for most practical purposes.}. The other external points must be kept fixed. If the lines coming out of $i$ and $i+r$ intersect then give a name to identify the intersection point, e.g. $R$ (If the lines do not intersect then go back to the original graph and move on to the next set of both $\alpha$- and $\beta$-consecutive external points).

Assuming the lines intersect, note that $\{ i,i+1,\ldots ,i+r, R\}$ form a convex polygon. Now remove the polygon from the graph, bring $R$ to the boundary of the disk and treat the new graph, which has $R$ as an external point, as a {\it new} problem and repeat the procedure. If at any given point one fails to find a set which forms a polygon with at most one internal point after trying all of them, then $m^{(0)}_n(\alpha|\beta)=0$.

This procedure comes to an end when one finds a graph where both orderings agree completely and therefore gives a single polygon with only `external' edges.

Finally, after completing all iterations and obtaining a list of all polygons found in the process one computes for each polygon its corresponding sub-amplitude. These sub-amplitudes can be computed just as regular amplitudes, where both orderings coincide, because for scalar particles it is straightforward to go off-shell.

The double partial amplitude, $m^{(0)}_n(\alpha|\beta)$, is then given by the product of all sub-amplitudes times a propagator for each internal point that was obtained in the process of removing polygons.

%\textbf{If there is at least one node that is not connected to any of its neighbors the intersection of empty and $m^{(0)}_n(\alpha|\beta)=0$. Assuming that every node is connected to at least one of its neighbors,} we can refine the diagram such that all consecutive nodes that are connected by a line are close enough to each other, see figure \ref{fig:example1a} (c). The external nodes must then belong to convex polygons with only one internal vertex. If this does not happen then the intersection is empty again. Each convex polygon can be thought of as a ``subamplitude" (with the internal node treated as a particle) given by a generating function with both permutations equal to the identity. Removing all polygons containing external particles we are left with a new graphs whose external nodes are the internal ones in the polygons that were removed. This new graph can be treated as before and the procedure should be repeated. If at any stage one meets the conditions for an empty intersection then one can stop and conclude that $m^{(0)}_n(\alpha|\beta)=0$. In our example, $m^{(0)}_8(I|54376218)$, the first iteration gives three convex polygons with vertices $\{ 2,1,8,B\}$, $\{ 3,4,5,A \}$, and $\{ 6,7,C\}$. The second iteration gives a single triangle with vertices $\{ A,B,C\}$.

\begin{figure}[h]
	\centering
		\includegraphics{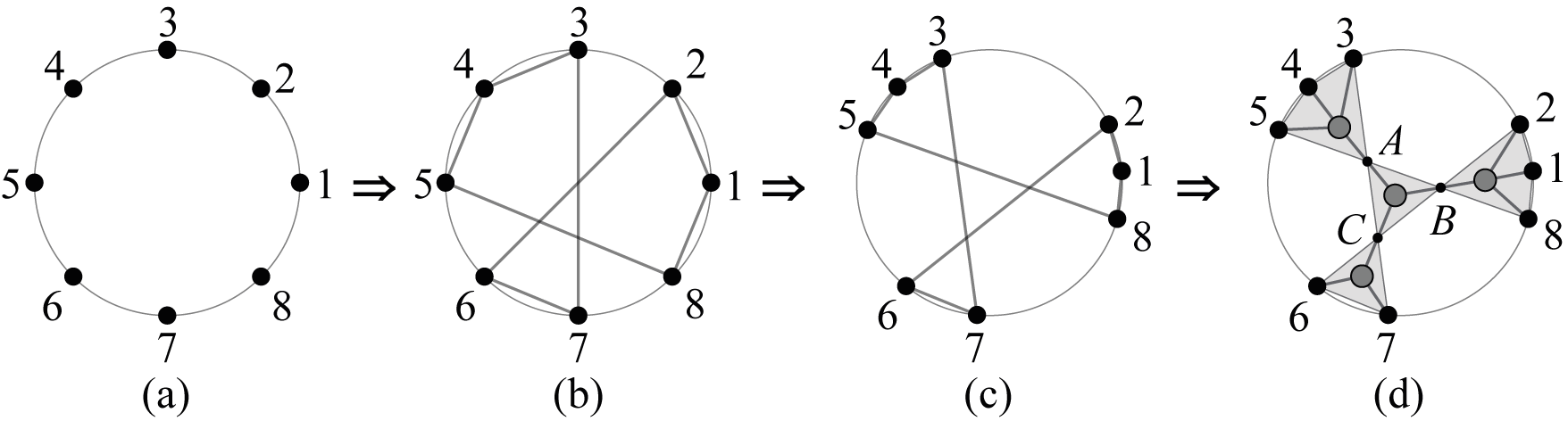}
	\caption{Computing $m^{(0)}_8(I|54376218)$ by finding its polygon decomposition. $(a)$ Points are drawn on the boundary of a disk according to the $\alpha$ ordering. $(b)$ A loop of line segments is drawn connecting the points according to the $\beta$ ordering. $(c)$ External points are moved along the boundary so that a polygon decomposition is manifest. In this example all polygons can be easily exhibited in a single step.}
	\label{fig:example1a}
\end{figure}

In figure \ref{fig:example1a} we present $m^{(0)}_8(I|54376218)$ as an illustrative example. In this particular case, by a simple rearrangement of the external points (from figure \ref{fig:example1a}(b) to figure \ref{fig:example1a}(c)) one finds a decomposition in terms of four convex polygons, namely, $\{ 2,1,8,B\}$, $\{ 3,4,5,A \}$, $\{ 6,7,C\}$ and $\{ A,B,C\}$. It is easy to obtain the corresponding sub-amplitudes (ignoring the signs)
%
%Finally, after completing all iterations and obtaining the set of all polygons one computes for each polygon its corresponding subamplitude. This is well defined, even when the polygon has internal vertices since scalar diagrams can be treated off-shell. The answer is the obtained by multiplying all subamplitudes and including a propagator for each internal vertex. In our example of has the following subamplitudes
%
\be
m^{(0)}(1,2,B,8|B,2,1,8) = \frac{1}{s_{21}}+\frac{1}{s_{18}},\quad m^{(0)}(3,4,5,A|5,4,3,A) = \frac{1}{s_{34}}+\frac{1}{s_{45}},
\nonumber
\ee
\be
m^{(0)}(6,7,C|7,6,C) =1,\quad m^{(0)}(B,A,C|B,A,C) =1,
\ee
and propagators
\be
A \rightarrow \frac{1}{s_{345}},\quad B\rightarrow \frac{1}{s_{812}},\quad C\rightarrow \frac{1}{s_{67}}.
\ee
Putting all of them together gives, up to an overall sign,
\be
m^{(0)}_8(I|54376218) = \left(\frac{1}{s_{21}}+\frac{1}{s_{18}}\right)\left(\frac{1}{s_{34}}+\frac{1}{s_{45}}\right)\frac{1}{s_{345}s_{812}s_{67}}.
\label{aso}\ee
%

%possibly after tuning the positions of the nodes (still respecting the ordering $\alpha$, Figure \ref{fig:example1a} (c)), if the graph is equivalent to a ``tree'' obtained by gluing a set of convex polygons upon their vertices, we replace each polygon by the set of all diagrams consistent with the number and the ordering of the polygon's vertices (Figure \ref{fig:example1a} (d), where this set is represented by a grey blob), and replace each gluing vertex by a scalar propagator. Gluing these ingredients together produces a set of $n$-point diagrams, whose summation is exactly the corresponding generating function (up to a sign).
%Especially, when $\beta$ differs from $\alpha$ at most by some cyclic permutation and flipping, such as that in \eqref{5ptcanonical}, the graph itself is already a convex $n$-gon, and so $m^{(0)}_n(\alpha|\beta)$ generates all diagrams consistent with the ordering $\alpha$ (or equivalently $\beta$).

%In our example $m^{(0)}_8(I|54376218)$, the graph can be determined as in Figure \ref{fig:example1a}.

%Here the three intersection points $A,B,C$ correspond to the propagators $s^{-1}_{345},s^{-1}_{128},s^{-1}_{67}$ respectively. The graph decomposes into two squares and two triangles. Since each triangle gives one $3$-pt planar diagram and each square gives two $4$-pt planar diagram with the prescribed ordering, $m^{(0)}_8(I|54376218)$ generates altogether four $8$-pt diagrams, whose weights are
%\be\label{examplediagrams}
%\frac{1}{s_{12}s_{128}s_{34}s_{345}s_{67}},\quad
%\frac{1}{s_{12}s_{128}s_{45}s_{345}s_{67}},\quad
%\frac{1}{s_{18}s_{128}s_{34}s_{345}s_{67}},\quad
%\frac{1}{s_{18}s_{128}s_{45}s_{345}s_{67}}.
%\ee

\begin{figure}[htb]
	\centering
		\includegraphics{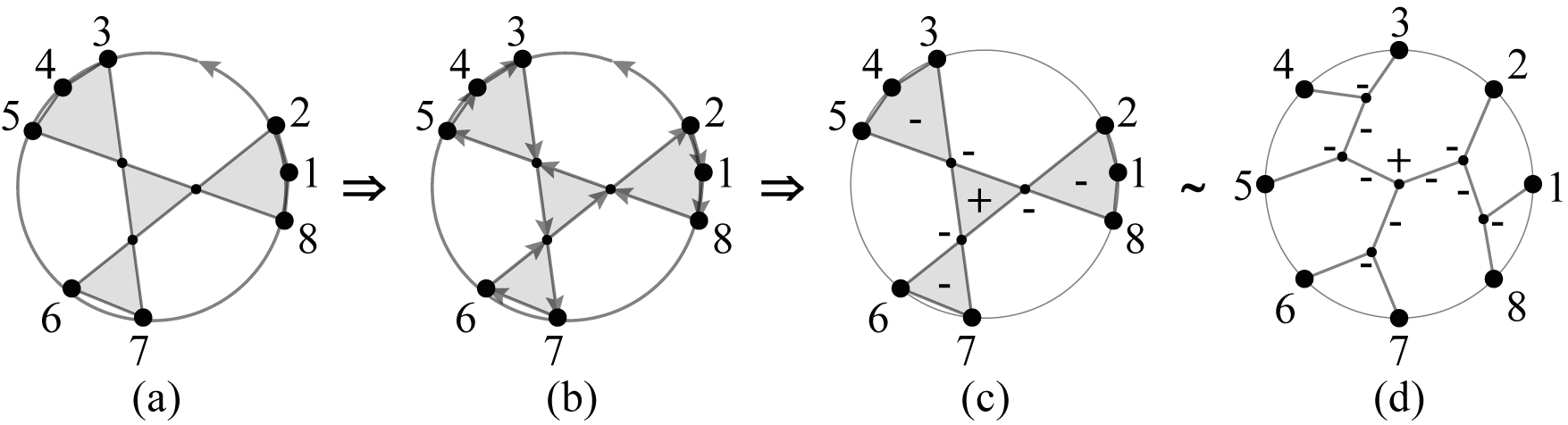}
	\caption{Sign of $m^{(0)}_8(I|54376218)$}
	\label{fig:example1b}
\end{figure}

Now we give the rule to determine the overall sign of the results. First define the orientation of the disk by the ordering $\alpha$ (Figure \ref{fig:example1b} (a)), and define the orientation of the loop of segments by the ordering $\beta$ (Figure \ref{fig:example1b} (b)), which induces an orientation in every convex polygon. The rule is as follows: (1) each polygon with odd number of vertices contributes a plus sign if its orientation is the same as that of the disk and a minus sign if opposite, (2) each polygon with even number of vertices always contributes a minus sign, and (3) each intersection point contributes a minus sign (Figure \ref{fig:example1b} (c)). Then the product of all these signs determines the overall sign of the double-partial amplitude relative to its corresponding scalar diagrams.

In our example one finds that the three external polygons give a minus sign while the internal one gives a plus sign. There are three propagators, and each gives a minus sign (see figure \ref{fig:example1b}). Altogether one finds six minus signs and therefore the overall sign is plus. This means that \eqref{aso} is indeed the right answer.

%And so for our example, as shown in Figure \ref{fig:example1b}, there are altogether $6$ minus sign, hence the overall sign is plus and we conclude that
%\be
%m^{(0)}_8(I|54376218)=\left(\frac{1}{s_{12}}+\frac{1}{s_{18}}\right)\frac{1}{s_{128}}\left(\frac{1}{s_{34}}+\frac{1}{s_{45}}\right)\frac{1}{s_{345}s_{67}}.
%\ee

Alternatively, one can write down any diagram from the double-partial amplitude (e.g.~Figure \ref{fig:example1b} (d)), and think of each cubic vertex as a triangle whose vertices are glued either to a node or to a vertex of another triangle. Then the sign can be obtained by the same rules as above. This is because in replacing each convex polygon by a specific diagram, the number of triangles and propagators are fixed by the polygon, and the triangles therein inherit the same orientation. Whenever a triangle picks up a minus sign, the ordering of its three vertices are flipped in $\beta$ as compared to $\alpha$~\footnote{When a vertex is not directly glued to a node, it is regarded as the set of all nodes that are linked to it indirectly. So it is easy to see that each triangle (or cubic vertex) induces a partition of the nodes into three sets consistent with both orderings $\alpha$ and $\beta$.}. In this way, the sign of the double-partial amplitude can be interpreted as the number of propagators $n-3$ and the number of ordering flips $n_{\text{flip}}(\alpha|\beta)$ in the cubic vertices of any specific diagram contained therein. This explains the factor
\be\label{overallsign}
(-1)^{n-3+n_{\text{flip}}(\alpha|\beta)}
\ee
introduced in \eqref{sumscalardiagrams}.

%assign a minus sign to each propagator, and assign to each cubic vertex if the three nodes it connects to are of the same ordering in both $\alpha$ and $\beta$ or a minus sign otherwise (when a leg does not connect to any labels directly, we regard it as the collection of all labels it connects to indirectly, which are always adjacent in both orderings). Then it is easy to see that the product of the signs gives the same answer (Figure \ref{fig:example1b} (d)), since if we regard each cubic vertex as a triangle then this is just a refinement of the tree of polygons in Figure \ref{fig:example1a} (d).

%In the other situation, if the graph can in no way be tuned into a tree of convex polygons, we always have $m^{(0)}_n(\alpha|\beta)=0$. Some examples of this type are $m^{(0)}_8(I|52786431)$, $m^{(0)}_8(I|18736524)$ and $m^{(0)}_8(I|36174852)$, as one may easily check by drawing the graphs according to the procedure described above.

\subsection{Relation to the KLT matrix}

We have shown that the formula for double-partial amplitudes $m^{(0)}(\alpha|\beta)$, \eqref{scal}, is a generating function of sums of trivalent diagrams for any given pair of permutations $\alpha,\beta$. Now we will give it another physical interpretation. Recall that in the KLT relations one defines the momentum kernel
\be  S[\alpha|\beta]=
\prod^{n{-}2}_{i=2}\left(s_{1, \alpha(i)}+\sum^{i{-}1}_{j=2} \theta(\alpha(j), \alpha(i))_{\beta} s_{\alpha(j),\alpha(i)}\right),
\ee
where $\alpha,\beta\in S_{n-3}$ are permutations acting on labels $2,3,\ldots,n{-}2$; $\theta(i,j)_\beta=1$ if the ordering of $i,j$ is the same in both sequences of labels, $\alpha(2),\ldots,\alpha(n{-}2)$ and $\beta(2),\ldots,\beta(n{-}2)$, and zero otherwise\footnote{The convention we use here follows that in~\cite{BjerrumBohr:2010ta}, where particle $1$ was chosen as a pivot.}.

We define the KLT matrix, $S_{\rm KLT}$ as a $(n{-}3)!$ by $(n{-}3)!$ matrix whose rows (superscript) and columns (subscript) are labeled by orderings $\alpha\equiv (1,\alpha(2),\ldots,\alpha(n{-}2),n{-}1,n)$ and $\beta\equiv (1,\beta(2),\ldots,\beta(n{-}2), n,n{-}1)$ (we use the same labels, $\alpha,\beta$ to denote the orderings and the permutations), and the entries are given by $( S_{\rm KLT})^\alpha_\beta=S[\alpha|\beta]$. 

At first sight, this matrix has nothing to do with double-partial amplitudes. However, here we will show that the inverse of the KLT matrix, is precisely given by a matrix, whose entries are double-partial amplitudes with corresponding pairs of orderings, $(m_{\rm scalar})^\alpha_\beta=m^{(0)}(1,\alpha(2),\ldots,\alpha(n{-}2),n{-}1,n|1,\beta(2),\ldots,\beta(n{-}2),n,n{-}1)$. The proof directly follows from a remarkable property of scattering equations called KLT orthogonality~\cite{Cachazo:2012da,Cachazo:2013gna}.

In order to state what KLT orthogonality is let us define a function defined for any two given solutions to the scattering equations $\{\sigma^{I}\}$ and $\{\sigma^{J}\}$
\be
(I,J) \equiv \sum_{\alpha,\beta\in S_{n-3}}V^{(I)}_{\alpha}S[\alpha|\beta]U^{(J)}_{\beta},\label{innerproduct}
\ee
where $V^{(I)}_{\alpha}$ stands for
\be
V_{\alpha} = \frac{1}{(\sigma_1-\sigma_{\alpha_2})(\sigma_{\alpha_2}-\sigma_{\alpha_3})\cdots (\sigma_{\alpha_{n-2}}-\sigma_{n-1})(\sigma_{n-1}-\sigma_{n})(\sigma_{n}-\sigma_{1})}~\label{V}
\ee
evaluated on the $I^{\rm th}$ solution while $U^{(J)}_{\beta}$ stands for the following expression evaluated on the $J^{\rm th}$ solution.
\be
U_{\beta} = \frac{1}{(\sigma_1-\sigma_{\beta_2})(\sigma_{\beta_2}-\sigma_{\beta_3})\cdots (\sigma_{\beta_{n-2}}-\sigma_{n})(\sigma_{n}-\sigma_{n-1})(\sigma_{n-1}-\sigma_{1})}~\label{U}.
\ee
As proven in \cite{Cachazo:2013gna}, the following holds for any $I$ and $J$,
\be
\frac{(I,J)}{(I,I)^{\frac 12}(J,J)^{\frac 12}} = \delta_{IJ}.
\label{KLT}\ee
This is known as KLT orthogonality.

For our purposes, only two facts, also explained in detail in \cite{Cachazo:2013gna}, are necessary. The first is that $S[\alpha|\beta]$ is only a function of the kinematic invariants $s_{ab}$. The second is that
\be
(J,J) = {\rm det}'\Phi(\sigma^{J}).
\label{que}\ee

It is interesting to note that the dimension of space of permutations of labels $2,\ldots,n{-}2$ (for which both $U$ and $V$ vectors form a basis, evaluated on any solution), and the dimension of the solution space, are both $(n{-}3)!$. This allows us to define the following $(n{-}3)!\times(n{-}3)!$ matrices based on $U,V$ vectors normalized with respect to the inner product \eqref{innerproduct}
\be
(\hat U)^I_\alpha\equiv \frac{U^{(I)}_\alpha}{(I,I)^{\frac 1 2}}, \quad (\hat V)^I_\beta\equiv \frac{V^{(I)}_\beta}{(I,I)^{\frac 1 2}},
\ee
where the rows (superscripts) are labeled by solutions and columns (subscripts) by permutations (or orderings), and KLT orthogonality is the simple statement that the product of these three matrices gives the identity matrix $\mathds{I}$ in solution space,
\be
\hat U S_{\rm KLT} \hat V^T=\mathds{I}.
\ee
It is obvious that $\hat V$ is invertible, thus by multiplying $\hat V^T$ and $(\hat V^T)^{-1}$ from left and right respectively, we obtain the identity matrix in permutation space,
\be
\mathds{I}=\hat V^T \hat U S_{\rm KLT} \hat V^T (\hat V^T)^{-1}={\hat V}^T \hat U S_{\rm KLT} \Leftrightarrow S_{\rm KLT}^{-1}=\hat V^T \hat U,
\ee
where by \eqref{objectsdef} we find the right-hand-side of the second equality is precisely given $m_{\rm scalar}$
\be
(S_{\rm KLT}^{-1})^\alpha_\beta=\sum_{I=1}^{(n{-}3)!} \frac{V^{(I)}_\alpha U^{(I)}_\beta}{{\rm det}' \Phi^{(I)}}=(m_{\rm scalar})^\alpha_\beta.
\ee

The inverse of the KLT matrix has also been discussed in~\cite{Broedel:2013tta} where it was related to the field-theory limit of string disk integrals\footnote{We thank Oliver Schlotterer for pointing out~\cite{Broedel:2013tta} to us, which motivated us to find the relation between double-partial amplitudes and the KLT matrix.}, following computations in~\cite{Mafra:2011nv,Mafra:2011nw}. As explicit examples, the inverse was given in~\cite{Broedel:2013tta} for up to seven points, and the result agrees with that of the double-partial amplitudes. It would be interesting to explore the connections further.

\section{Color-Kinematics Duality}

In the introduction we illustrated how our formulation relates scalar-, gluon- and graviton-amplitudes by simple transformations ($C\rightarrow E$ or $\tilde{C}\rightarrow\tilde{E}$ or both). This replacement occurs solution by solution of the scattering equations. More explicitly,
\be
{\cal M}^{(0)}_n =\!\!\sum_{I=1}^{(n-3)!}\frac{C(\sigma^{(I)}){\tilde C}(\sigma^{(I)})}{{\rm det}'\Phi(\sigma^{(I)})},\quad {\cal M}^{(1)}_n = \!\!\sum_{I=1}^{(n-3)!}\frac{C(\sigma^{(I)}){\tilde E}(\sigma^{(I)})}{{\rm det}'\Phi(\sigma^{(I)})},\quad {\cal M}^{(2)}_n = \!\!\sum_{I=1}^{(n-3)!}\frac{E(\sigma^{(I)}){\tilde E}(\sigma^{(I)})}{{\rm det}'\Phi(\sigma^{(I)})}.
\ee
Very interestingly, in \cite{Hodges:2011wm} Hodges made the observation that using twistor diagrams the statement that ``gravity is the squared of Yang-Mills" is replaced by ``gravity times $\phi^3$ is the square of Yang-Mills". Indeed, our construction shows that this is precisely true solution by solution of the scattering equations!

Also mentioned in the introduction is the fact that the color factor $C$ and the kinematic factor $E$ have very similar properties and this is the reason why they are exchangeable. It is very natural to suspect that this color-kinematics duality must have a connection to the color-kinematics duality introduced by Bern, Carrasco and Johansson (BCJ) \cite{Bern:2008qj}.

Indeed, in this section we show that these simple transformations are equivalent to the BCJ doubling procedure to obtain gravity amplitudes from Yang-Mills ones.

\subsection{Expansion of the Pfaffian}

Here we show that the Pfaffian can be expanded in a way similar to the expansion of the color factors. We choose to expand the color factor, $C$ ($\tilde{C}$), in terms of the basis given in \eqref{c-basis},
\be
C=\sum_{\gamma\in S_{n-2}}\frac{{\bf c}_{1\gamma(2)\cdots\gamma(n-1)n}}{\sigma_{1,\gamma(2)}\cdots\sigma_{\gamma(n-1),n}\sigma_{n,1}},
\label{caso}\ee
%where we use the abbreviation
%\be\label{colorexpansion}
%F_{1\gamma(2)\cdots\gamma(n-1)n}=f_{\textsf{a}_{1}\textsf{a}_{\gamma(2)}\textsf{b}_1}f_{\textsf{b}_1\textsf{a}_{\gamma(3)}\textsf{b}_2}\cdots f_{\textsf{b}_{n-3}\textsf{a}_{\gamma(n-1)}\textsf{a}_n},
%\ee
and a similar formula holds for $\tilde{C}$.

This form of $C$ hints that a similar form for $E$ should exist. More explicitly, there must exist functions, denoted as ${\bf n}$, which only depend on kinematic data $\{\epsilon^{\mu}_a,k^\mu_a \}$, such that
\be
E = {\rm Pf}'\Psi(\epsilon,k,\sigma ) = \sum_{\gamma\in S_{n-2}}\frac{{\bf n}_{1\gamma(2)\cdots\gamma(n-1)n}}{\sigma_{1,\gamma(2)}\cdots\sigma_{\gamma(n-1),n}\sigma_{n,1}}.
\label{hope}\ee
Straightforwardly expanding the reduced Pfaffian ${\rm Pf}'\Psi(\epsilon,k,\sigma )$ leads to an expression very different from the right hand side. In fact, at first sight it seems difficult to rewrite it into the form of \eqref{hope}.

Luckily, not only \eqref{hope} holds but the proof is very simple again thanks to the KLT orthogonality~\cite{Cachazo:2012da,Cachazo:2013gna}. In order to prove that the expansion in \eqref{hope} exists it is enough to rewrite \eqref{KLT} as (recall the definitions in~\eqref{V},\eqref{U})
\be
\frac{(I,J)}{(J,J)} = \delta_{IJ},
\ee
then to multiply it by ${\rm Pf}'\Psi(\sigma^J)$ and sum over $J$ get
\be
\sum_{J=1}^{(n-3)!}\frac{(I,J){\rm Pf}'\Psi(\sigma^J)}{(J,J)} = {\rm Pf}'\Psi(\sigma^I).
\label{jum}\ee
Using %the explicit form of $(I,J)$
\eqref{innerproduct} and \eqref{que} one finds that the left hand side of \eqref{jum} is given by
\be
\sum_{\alpha,\beta\in S_{n-3}}V^{(I)}_{\alpha}S[\alpha|\beta]\sum_{J=1}^{(n-3)!} \frac{U^{(J)}_{\beta}{\rm Pf}'\Psi(\sigma^J)}{{\rm det}'\Phi(\sigma^{J})},
\ee
where note that $\alpha$ and $\beta$ are permutations of labels $2,3,\ldots,n{-}2$. The sum over $(n-3)!$ solutions appearing in this equation is nothing but the Yang-Mills partial amplitude $M^{(1)}_n(1,\beta ,n,n-1)$. Therefore, on the support of scattering equations,
\be
{\rm Pf}'\Psi(\sigma) = \sum_{\alpha\in S_{n-3}}\frac{\sum_{\beta\in S_{n-3}}S[\alpha|\beta]M^{(1)}_n(1,\beta ,n,n-1)}{(\sigma_1-\sigma_{\alpha_2})(\sigma_{\alpha_2}-\sigma_{\alpha_3})\cdots (\sigma_{\alpha_{n-2}}-\sigma_{n-1})(\sigma_{n-1}-\sigma_{n})(\sigma_{n}-\sigma_{1})}.
\ee
This concludes the proof of \eqref{hope} since we have found an explicit form of the numerators in terms of only external kinematic invariants. More explicitly,
\be
{\bf n}_{1\gamma(2)\cdots\gamma(n-1)n} = \begin{cases}  \displaystyle \sum_{\beta\in S_{n-3}}S[\gamma|\beta]M^{(1)}_n(1,\beta ,n,n-1), \quad \gamma(n-1) = n-1,\\ \displaystyle 0,\quad \gamma(n-1)\neq n-1.\end{cases}\label{BCJnumerator}
\ee
The form given in this proof (which was also discussed in e.g.~\cite{Kiermaier10talk}) is not unique and not very useful for practical computations as it is tautological in nature. However, it shows that expansions of the form \eqref{hope} {\it do} exists if we regard the $M^{(1)}_n$'s in \eqref{BCJnumerator} as functions of external data, and this is all we need in order to make the connection to the BCJ doubling construction. As we will see shortly, the $(n{-}2)!$ ${\bf n}$'s play the role of a basis for BCJ numerators, and in practice it is always possible to derive local expressions for them by carefully using scattering equations when expanding $E$; explicit expressions for such BCJ numerators can be found in~\cite{Mafra:2011kj,Fu:2012uy}.
%On the other hand, it is not hard to observe that after we expand the reduced Pfaffian in $E$ ($\tilde{E}$), in the denominator of each term every $\sigma$ variable appears exactly twice. Here we leave as a claim that by properly using scattering equations and identities like $\frac{1}{\sigma_{ab}\sigma_{bc}}=\frac{1}{\sigma_{ac}\sigma_{ab}}+\frac{1}{\sigma_{ac}\sigma_{bc}}$, the expansion of $E$ can be transformed also into the form
%\be\label{kinematicsexpansion}
%E=\sum_{\gamma\in S_{n-2}}\frac{N_{1\gamma(2)\cdots\gamma(n-1)n}}{\sigma_{1,\gamma(2)}\cdots\sigma_{\gamma(n-1),n}\sigma_{n,1}},
%\ee
%where $N_{1\gamma(2)\cdots\gamma(n-1)n}$ is a function only of the kinematics data, the explicit form of which depends on specific transformations.

Now we can unify \eqref{caso} and \eqref{hope} by denoting both ${\bf c}$ and ${\bf n}$ as $e$ in all three theories. Then if we expand the product of the two summations, each term is of the form
\be\label{singleterm}
e_{1\alpha(2)\cdots\alpha(n-1)n}\tilde{e}_{1\beta(2)\cdots\beta(n-1)n}m^{(0)}_n(1\alpha(2)\cdots\alpha(n-1)n|1\beta(2)\cdots\beta(n-1)n)\equiv e_\alpha \tilde{e}_\beta m^{(0)}(\alpha|\beta).
\ee

The full amplitude for scalar, pure Yang-Mills or gravity can be written in a unified form,
\be
{\cal M}^{({\bf s})}_n=(-1)^{n{-}3}\sum_{\alpha,\beta\in S_{n{-}2}} (-)^{n_\text{flip}(\alpha|\beta)} e_\alpha \tilde{e}_\beta \!\!\sum_{g\in \mathcal{T}(\alpha)\cap\mathcal{T}(\beta)}\prod_{e\in E(g)} \frac 1{s_e},\label{unified_ck}
\ee
where for $e,\tilde e$ we have ${\bf s}$ kinematic numerators and $2{-}{\bf s}$ color numerators, with ${\bf s}=0,1,2$. In this form the color-kinematics correspondence of our formula becomes more transparent: the $\sigma$-independent color and kinematic factors ${\bf c}_\alpha$ and ${\bf n}_\alpha$ are on a equal footing, and by exchanging them one relates amplitudes of scalar theory, Yang-Mills and gravity.

\subsection{Relation to BCJ Color-Kinematics Duality}

To see that \eqref{unified_ck} actually gives a representation which respects the BCJ color-kinematics duality, we can start by exchanging the two summations. First note that the union of all sets $\mathcal{T}(\alpha)\cap\mathcal{T}(\beta)$ with $\alpha,\beta\in S_{n{-}2}$ is the complete set of trivalent diagrams with $n$ legs, ${\cal T}_n$. For each trivalent diagram $g$, we define the set of {\it pairs of orderings}\footnote{Here we slightly abuse the notation, by an ordering $\alpha$ we mean the sequence $(1,\alpha(2),\ldots,\alpha(n{-}1),n)$ obtained from the permutation $\alpha$.} that can generate $g$, $PO(g)$, and it is obvious that $(\alpha,\beta)\in PO(g)\Leftrightarrow (\beta,\alpha)\in PO(g)$. Thus $PO(g)=O(g)\otimes O(g)$ which defines $O(g)$: $g \in \mathcal{T}(\alpha)\cap\mathcal{T}(\beta)\Leftrightarrow \alpha,\beta\in O(g)$. The formula can be written as a sum of all cubic diagrams, each with propagators and two numerators,
\be
{\cal M}^{({\bf s})}_n=(-1)^{n{-}3}\sum_{g \in {\cal T}_n} \prod_{e\in E(g)} \frac 1{s_e} \sum_{(\alpha,\beta)\in O(g)\otimes O(g)} (-)^{n_\text{flip}(\alpha|\beta)} e_\alpha \tilde{e}_\beta.
\ee

Note that the signs satisfy a composition identity, which simply follow from its definition in the previous section: for any $\gamma\in O(g)$, $(-1)^{n_\text{flip}(\alpha|\beta)}=(-1)^{n_\text{flip}(\alpha|\gamma){+}n_\text{flip}(\gamma|\beta)}$, thus the double sum over $\alpha,\beta$ factorizes into two sums,
\be
\sum_{(\alpha,\beta)\in O(g)\otimes O(g)} (-)^{n_\text{flip}(\alpha|\beta)} e_\alpha \tilde{e}_\beta=\left(\sum_{\alpha\in O(g)} (-1)^{n_\text{flip}(\alpha|\gamma)} e_\alpha \right) \left(\sum_{\beta\in O(g)} (-1)^{n_\text{flip}(\gamma|\beta)} \tilde{e}_\beta\right)\equiv e_g \tilde e_g,\label{defofnum}
\ee
where we have defined $e_g$ and $\tilde e_g$ to be the combinations in the two brackets respectively. Note that this definition depends on the choice of $\gamma$ that is fixed for the sums. Different choices of $\gamma$ may leads to different signs, and when $\{e_g\}$ corresponds to color factors, this is the standard sign ambiguity for generic trivalent diagrams (in the special case when $e_g$ receives contributions from only one $e_{\alpha}$, we can choose $\gamma=\alpha$).
%In the special case of the multi-peripheral diagrams naturally associated with an ordering $\alpha$, denoted as $g_\alpha$, we can choose $\gamma=\alpha$ such that $e_{g_\alpha}=e_\alpha$ agrees with the standard convention.

\begin{figure}[h]
\centering
\includegraphics{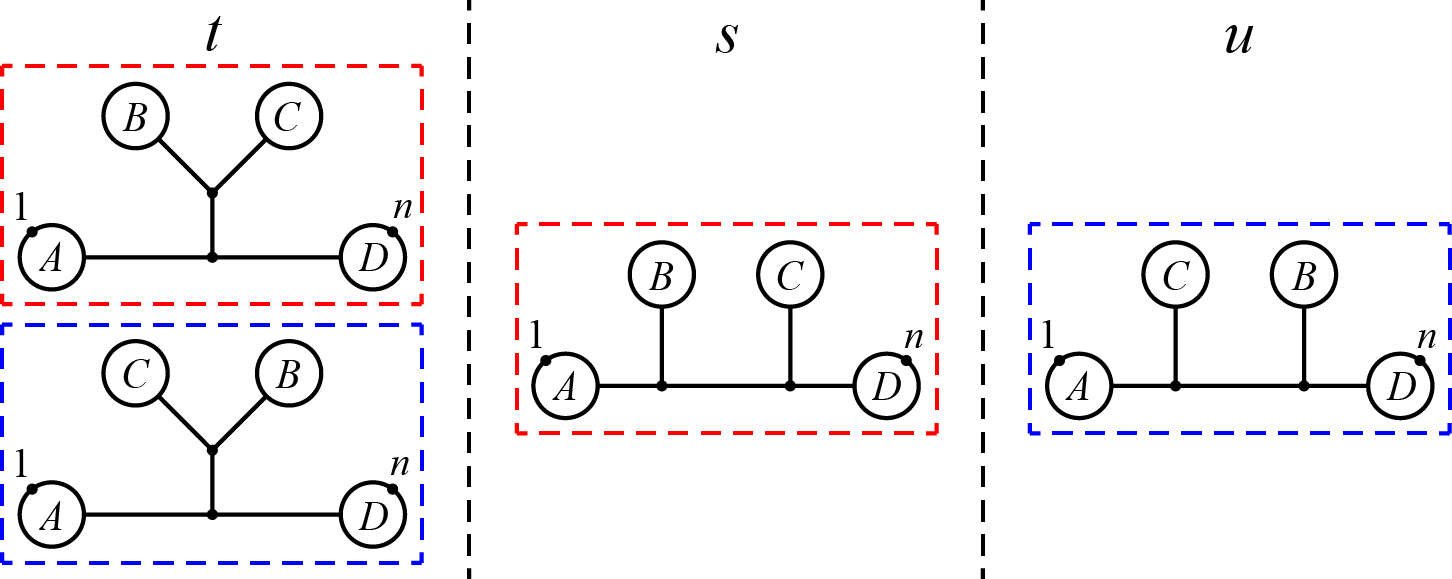}
\caption{Trivalent diagrams $g_t,g_s,g_u$ when particle $1,n$ are contained in two different trees, e.g. $A, D$, attached to the four-particle subdiagram. Red and blue regions correspond to $O(g_s)$ and $O(g_u)$ respectively, the union of which gives $O(g_t)$. }
\label{orderings1}
\end{figure}

In this way, up to possible signs, our formula defines numerator factors $e_g, \tilde e_g$ for every trivalent diagram, and the physical quantity which is the product of the two copies, $e_g \tilde e_g$, has no such ambiguities. We now argue that these numerators automatically satisfy Jacobi-like identities.
Given three trivalent trees which differ only by a four-particle subdiagram, as illustrated in figure.~\ref{orderings1}, we need to prove that the numerators $e_g$ (similarly for $\tilde e_g$) satisfy a Jacobi identity,
\be
e_{g_t}=\pm(e_{g_s}-e_{g_u}),
\ee
and here we emphasize that the relative sign of the two terms on the right-hand-side is important.

Let us denote the four trees attached to the four legs of the subdiagram as $A,B,C$ and $D$, and we first consider the simple case where particles $1, n$ are attached to two different trees, which, without loss of generality, we assume to be $A, D$ respectively (see figure.~\ref{orderings1}). The idea is that the set of orderings for the complete diagram can be obtained from putting together the four sets corresponding to the four trees. For the tree $B (C)$, we remove the label of the internal leg from each $\alpha\in O(B)$ ($\alpha \in O(C)$), and define the new set which only has labels for external particles as $O^*(B)$ $(O^*(C))$; we do the same thing for $A$ and $D$, but in the definition of $O^*(A)$ we only keep the orderings with $1$ in one end, and in $O^*(D)$ only those with $n$ in the other. It is straightforward to find that,
\be
O(g_s)=\{(\alpha_A,\alpha_B,\alpha_C,\alpha_D)\},\,O(g_u)=\{(\alpha_A,\alpha_B,\alpha_C,\alpha_D)\},\,{\rm with}\,\alpha_i \in O^*(i)\,{\rm for}\,i=A,B,C,D,
\ee
while $O(g_t)=O(g_s)\bigcup O(g_u)$, and the three sets have been indicated in Figure~\ref{orderings1}. Note that in the three sets $1, n$ are on the two ends of the orderings, and it is trivial to see that $O(g_s)\bigcap O(g_u)=\varnothing$. Using~\eqref{defofnum} we have
\ba e_{g_t}&=&\sum_{\alpha\in O(g_t)}(-1)^{n_\text{flip}(\alpha|\gamma_t)} e_\alpha\nl
&=&(-1)^{n_\text{flip}(\gamma_s|\gamma_t)}\sum_{\alpha\in O(g_s)}(-1)^{n_\text{flip}(\alpha|\gamma_s)} e_\alpha+(-1)^{n_\text{flip}(\gamma_u|\gamma_t)} \sum_{\alpha\in O(g_u)}(-1)^{n_\text{flip}(\alpha|\gamma_u)}e_\alpha,\nonumber
\ea
where $\gamma_{t,s,u}\in O(g_{t,s,u})$ are the elements chosen in the definition of $e_{g_{t,s,u}}$, and in the second line the two summations give $e_{g_s}$ and $e_{g_u}$ respectively. The sign for an individual $e$ does not matter, but the relative sign of the two terms, $(-1)^{n_\text{flip}(\gamma_s|\gamma_u)}$, is unambiguous: from our definition of $n_\text{flip}$, this sign is determined by looking at the relative orientation for each cubic vertex of $g_t$ in the ordering $\gamma_s$ and $\gamma_u$, and the only difference appears in the cubic vertex with three internal legs, where $B$ and $C$ are exchanged from $\gamma_s$ to $\gamma_u$, thus the relative sign of $e_{g_s}$ and $e_{g_u}$ is always $-1$.

\begin{figure}[h]
\centering
\includegraphics{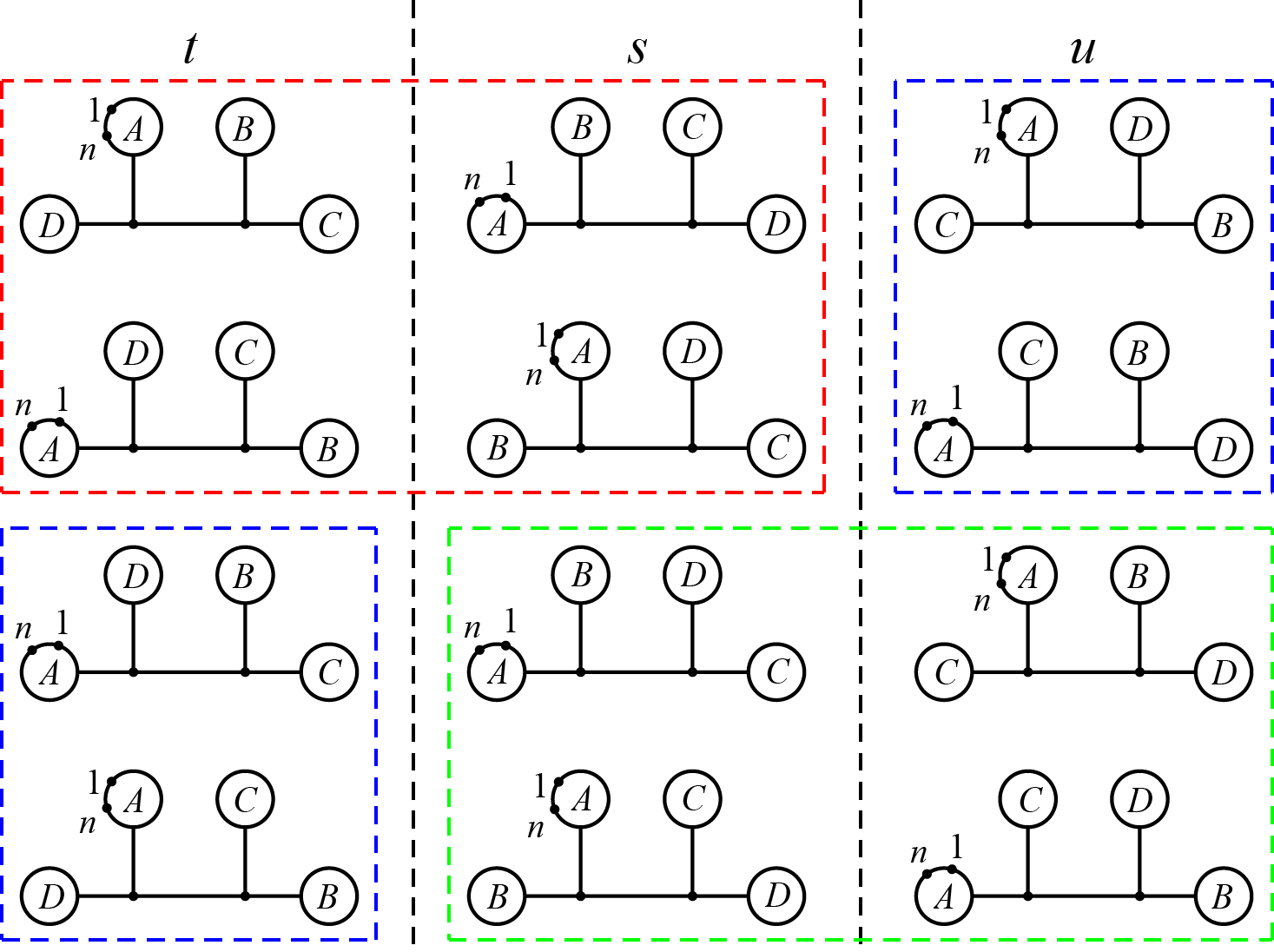}
\caption{Trivalent diagrams $g_t,g_s,g_u$ when particle $1,n$ are contained in a single tree, e.g. $A$, attached to the four-particle subdiagram. Green, red and blue regions correspond to $O=O(g_s)\bigcap O(g_u)$, $O(g_s)/O$ and $O(g_u)/O$ respectively. It is easy to see that $O(g_t)=O(g_s)\bigcup O(g_u)/O$ is the union of red and blue regions.}
\label{orderings2}
\end{figure}

The case when particle $1,n$ are attached to the same tree, e.g. $A$ , see figure.~\ref{orderings2}, can be argued similarly. To keep the expressions short, we denote the ordering corresponding to tree $A$ by two groups of ellipses which contain $1$ and $n$ respectively. The sets of orderings for $g_s,g_t,g_u$ can be written as
\ba
&& O(g_s)=\{(\ldots,\{\alpha_B,\{\alpha_C,\alpha_D\}\},\ldots)\},\,O(g_u)=\{(\ldots,\{\alpha_C,\{\alpha_B,\alpha_D\}\},\ldots)\},\\
&& O(g_t)=\{(\ldots,\{\alpha_D,\{\alpha_B,\alpha_C\}\},\ldots)\},\,{\rm with}\,\alpha_i \in O^*(i)\,{\rm for}\,i=B,C,D,
\ea
where $\{\alpha,\beta\}$ means that the two objects are unordered, and there are four types of elements in each set depending on how $B,C,D$ are ordered, as illustrated in figure.~\ref{orderings2}. Note that in this case $O\equiv O(g_s) \bigcap O(g_u)\neq \varnothing$ (the green region in figure.~\ref{orderings2}), which has elements of the form $(\ldots,\alpha_B,\alpha_D,\alpha_C,\ldots)$ or $(\ldots,\alpha_C,\alpha_D,\alpha_B,\ldots)$. The two sets $O(g_s)/O$ and $O(g_u)/O$ are indicated by the red and blue regions in figure.~\ref{orderings2}, and it is obvious that their union gives $O(g_t)$. 
%$O_{g_t}\bigcap O=\emptyset$ and $O_{g_s}\bigcup O_{g_u}/O=O(g_t)$.
Again from~\eqref{defofnum}, up to a possible overall sign, the difference of $e_{g_s}$ and $e_{g_u}$ is given by the following expression with $\gamma_s\in O(g_s)/O$, $\gamma_u\in O(g_u)/O$,
\ba
e_{g_s}-e_{g_u}&=&\sum_{\alpha\in O(g_s)/O}(-1)^{n_\text{flip}(\alpha|\gamma_s)} e_\alpha+ (-1)^{n_\text{flip}(\gamma|\gamma_s)}\sum_{\alpha\in O}(-1)^{n_\text{flip}(\alpha|\gamma)}e_\alpha \nl &&-(-1)^{n_\text{flip}(\gamma_u|\gamma_s)}\sum_{\alpha\in O(g_u)/O}(-1)^{n_\text{flip}(\alpha|\gamma_u)}e_\alpha-(-1)^{n_\text{flip}(\gamma|\gamma_u)}\sum_{\alpha\in O}(-1)^{n_\text{flip}(\alpha|\gamma)} e_\alpha,\nonumber
\ea
where $\gamma\in O$, and the three sign factors all give $-1$. Note that the second and the fourth term cancel with each other, and the first and the third term combine into, up to a sign, $e_{g_t}$, thus we again find $e_{g_t}=\pm (e_{g_s}-e_{g_u})$.

Since the above statement holds regardless of $e, \tilde e$ being color or kinematic numerators, it guarantees that the representation of $\mathcal{M}^{({\bf s})}_n$ as a sum of cubic diagrams given by our formula automatically respects color-kinematics duality.
%and the numerators for the multi-peripheral diagrams, $\{e_{g_\alpha}\}$, $\{\tilde{e}_{g_\beta}\}$ serve as the basis for each of the duality-respecting numerators. 
Moreover, the simple transformations between $C$ ($\tilde C$) and $E$ ($\tilde E$) in our formula correspond to the substitution of the set of $e$ factors into a set of $e'$ factors which satisfy the same algebraic relations (or $\tilde{e}$ into $\tilde{e}'$), thus they are equivalent to the double-copy procedures that relate scalar, Yang-Mills and gravity amplitudes. Proofs of the existence of dualtiy-respecting numerators and the double-copy relations can also be found in~\cite{Bern:2010yg} \cite{Mafra:2011kj}.

To conclude this section it is important to recall a simple and well-known property of formulas of the form
\be
{\cal M}^{({\bf s})}_n=(-1)^{n{-}3}\sum_{g \in {\cal T}_n} e_g{\tilde e}_g\prod_{e\in E(g)} \frac{1}{s_e}.
\ee
Given any three graphs, as the ones discussed above, $g_s,g_u,g_t$, which only differ in a single propagator, i.e.,
\be
s\prod_{e\in E(g_s)} \frac{1}{s_e} =t\prod_{e\in E(g_t)} \frac{1}{s_e} =u\prod_{e\in E(g_u)} \frac{1}{s_e} 
\ee
one can freely redefine a kinematic set of numerators, say, $\{ e_g\}$ according to 
\be
e_{g_s}\to e'_{g_s} = e_{g_s} + \omega s,~ e_{g_t}\to e'_{g_t} = e_{g_t} + \omega t, ~ e_{g_u}\to e'_{g_u} = e_{g_u} - \omega u,
\ee
and the value of the formula remains invariant. This is simply because the set $\{\tilde e_g\}$, which can be either color or kinematic numerators, satisfy Jacobi-like identities.

In general the new numerators $\{e'_g\}$ obtained by applying any number of transformations of this form, with completely arbitrary parameters $w$, will not satisfy Jacobi-like identities. Applying this to Yang-Mills amplitudes one can obtain formulas of the form
\be
{\cal M}^{(1)}_n = (-1)^{n{-}3}\sum_{g \in {\cal T}_n} {\bf c}_g {\bf n}'_g\prod_{e\in E(g)} \frac{1}{s_e}
\label{dao}\ee
with kinematic numerators $\{ {\bf n}'_g\}$ which do not satisfy Jacobi-like relations starting from   
\be
{\cal M}^{(1)}_n = (-1)^{n{-}3}\sum_{g \in {\cal T}_n} {\bf c}_g {\bf n}_g\prod_{e\in E(g)} \frac{1}{s_e}
\label{mii}\ee
where $\{ {\bf n}_g\}$ satisfy the Jacobi-like relations. 

Clearly, one can start with a gravity formula given by $e_g = {\bf n}_g$ and ${\tilde e}_g = {\bf n}_g$ and apply the same transformations that led to \eqref{dao} on the ${\tilde e}_g$ factors to get
\be
{\cal M}^{(2)}_n = (-1)^{n{-}3}\sum_{g \in {\cal T}_n} {\bf n}_g {\bf n}'_g\prod_{e\in E(g)} \frac{1}{s_e}.
\label{graf}\ee
This result can be interpreted as saying that the gravity formula \eqref{graf} can be obtained by using a double-copy procedure applied to \eqref{dao} and \eqref{mii} and therefore conclude that the double-copy procedure works even when one of the two Yang-Mills numerators do not satisfy Jacobi-like identities\footnote{We thank Yu-tin Huang for discussions that led to this point.}.

\section{Special Kinematics: Generating Catalan Numbers}

The Feynman diagrams of a colored cubic scalar theory are all possible trivalent, color-ordered planar trees. Each internal edge of the tree is dressed with a propagator factor and its contribution to the amplitude is the product of all propagators in the tree. In spacetime dimensions large enough compared to the number of particles, the set of kinematic invariants
\be
(k_i+k_{i+1}+\ldots +k_{i+r})^2 \quad {\rm with} \quad i\in \{1,2,\ldots ,n\},\, r\in \{2,3,\ldots ,\left[\frac{n}{2}\right]\},
\ee
modulo momentum conservation, form a basis of the $n(n-3)/2$ dimensional space of kinematic invariants. This means that we can choose any values for them which then completely specifies a single kinematic point.

In this section we consider the point where
\be
(k_i+k_{i+1}+\ldots +k_{i+r})^2 = 1
\ee
for all possible values of $i$ and $r$. The motivation for doing this is that each Feynman diagram contributes exactly $1$ to the amplitude. Therefore $m^{0}_n\equiv m^{(0}_n(I|I)$ with $I=(1,2,\ldots ,n)$ is simply the number of planar ordered trees. This number is known to be the Calatan number $C_{n-2}$, i.e.,
\be
m^{(0)}_n(1,2,\ldots ,n) =  \frac{(-1)^{n+1}}{n+1}\left(
                                           \begin{array}{c}
                                             2n \\
                                             n \\
                                           \end{array}
                                         \right).
\ee
At the special kinematic point it is easy to compute that all two particle kinematic invariants are
$\{s_{i,i+1} =1, s_{i,i+2}=-1\}$ for $i\in \{1,\ldots ,n\}$ (indexes are understood modulo $n$) and zero otherwise.

The scattering equations simplify dramatically and become
\be
-\frac{1}{\sigma_a-\sigma_{a-2}}+\frac{1}{\sigma_a-\sigma_{a-1}}+\frac{1}{\sigma_a-\sigma_{a+1}}-
\frac{1}{\sigma_a-\sigma_{a+2}} = 0 \quad {\rm for} \quad a\in \{1,2,\ldots, n\}.
\ee
The most direct way to solve the equations is to use $n{-}4$ equations to express $n{-}4$ $\sigma$'s in terms of the remaining one, and when plugging into the last equation, it becomes a polynomial equation of the remaining $\sigma$, which one can solve to get all the solutions. Here we follow a different route which turns out to be more instructive.  The equations are $\SL2C$ invariant, and one can rewrite them in terms of the cross-ratios:
\be
u_{i,j}\equiv \frac{(\sigma_i-\sigma_{j{+}1})(\sigma_{i{+}1}-\sigma_j)}{(\sigma_i-\sigma_j)(\sigma_{i{+}1}-\sigma_{j{+}1})},
\ee
where by definition $u_{i,i\pm 1}=0$ and $u_{i,i}=\infty$. It is straightforward to see that the $a^{\rm th}$ scattering equation at the special kinematic point becomes,
\be (u_{a{-}1,a{+}1}-u_{a{-}2,a})(\frac 1{\sigma_a-\sigma_{a{+}1}}-\frac 1{\sigma_a-\sigma_{a{-}1}})=0.\ee
Since the second factor does not vanish (unless $\sigma_{a{-}1}=\sigma_{a{+}1}$ which represents a unwanted, singular kinematic point), we conclude that the scattering equations are equivalent to
\be u_{1,3}=u_{2,4}=\ldots=u_{n,2},\label{creqs}\ee
which corresponds to a symmetric configuration with all cross-ratios built from four consecutive points equal to each other.

Therefore, in terms of the cross-ratios, the scattering equations become trivial. On the other hand, by definition the cross-ratios satisfy constraints which correspond to the so-called ``Y-system'' equations~\cite{Alday:2009dv}
\be
(1-u_{i, j{+}1})(1-u_{i{+}1,j})=(1-\frac 1{u_{i,j}})(1-\frac 1 {u_{i{+}1,j{+}1}}),\label{Yeqs}
\ee
and it turns out we can directly solve the combined system of \eqref{creqs} and \eqref{Yeqs}! Given the fact $u_{i,i{+}2}$ is independent of $i$, we immediately see that for $m=2,3,\ldots,n{-}2$ ($m=0,1,n{-}1,n$ are trivial), the cross-ratios $u_{i,i{+}m}$ are independent of $i$. Thus we can define $R_m\equiv u_{i,i{+}m}$, and by a simple change of variable $R_m=\frac {Y_m} {1+Y_m}$ the equations in terms of the new variables become precisely the Y-system equations in the high temperature limit considered in~\cite{Alday:2009dv},
\be
(1-R_{m{-}1})(1-R_{m{+}1})=(1-R_m)^2 \Leftrightarrow (1+Y_{m{-}1})(1+Y_{m{+}1})=Y_m^2,
\ee
with the boundary condition $Y_0=-1$, $Y_1=0$. It is well known that there are $\lfloor(n{-}1)/2\rfloor$ solutions, labeled by $i$, of the following form,
\be
Y^{(i)}_m=\frac{\sin(2\pi i(m{-}1)/n)\sin(2\pi i(m{+}1)/n )}{\sin^2(2 \pi i/n)},\label{Ysol}
\ee
for $i=1,\ldots,\lfloor(n{-}1)/2\rfloor$.

Based on the solutions, one can evaluate the double-partial amplitude at the special kinematic point,
\be
m^{(0)}_n(1,2,\ldots,n)=\sum_{i=1}^{\lfloor(n{-}1)/2\rfloor} \frac 1{(\sigma_{12}\ldots\sigma_{n1})^2\det'\Phi(Y_m^{(i)})},
\ee
where $...(Y_m^{(i)})$ means that the full integrand is a rational function of the $Y_m$ variables. For general $n$ and kinematic points, the integrand is a rather complicated function of cross-ratios; on the other hand, with the special solutions~\eqref{Ysol}, as we have checked up to $n=20$, the summand simplifies significantly, which we conjecture to be true for general $n$.
\vskip0.1in
{\it Conjecture}: At the special kinematic point, the formula as a function of $Y^{(i)}_m$ in \eqref{Ysol}  is
\be \frac 1{(\sigma_{12}\ldots\sigma_{n1})^2\det'\Phi(Y_m^{(i)})}=-\frac  {2^{n{-}1}} n (\cos (\frac{2\pi\hat{i}}{n})-1)^{n{-}2}(\cos (\frac{2\pi\hat{i}}{n})+1),\ee
where $\hat{i}=i$ for $n$ even, and $\hat{i}=i-\frac 1 2$ for $n$ odd.
\vskip0.1in

From this it is straightforward to see that the amplitude at the special kinematic point indeed counts the number of planar trivalent diagrams, which is the Catalan number.
\vskip0.1in
{\it Corollary}: The double-partial amplitude at the special kinematic point is given by
\be m^{(0)}_n(1,2,\ldots,n)=-\frac  {2^{n{-}1}} n \sum_{i=1}^{\lfloor(n{-}1)/2\rfloor} (\cos (\frac{2\pi\hat{i}}{n})-1)^{n{-}2}(\cos (\frac{2\pi\hat{i}}{n})+1)=\frac{(-1)^{n+1}}{n+1}\left(
                                           \begin{array}{c}
                                             2n \\
                                             n \\
                                           \end{array}
                                         \right), \label{Catalan}\ee

\vskip0.1in
{\it Proof}: Consider the Dynkin diagram of $\textbf{A}_{n{-}1}$, which consists of $n{-}1$ vertices and $n{-}2$ edges connecting them in sequence. By the combinatoric definition of Catalan number, $C_k$ is the number of closed paths with length $2k$ from one end of the diagram to itself. Denote the vertices as $1,2,\ldots,n{-}1$, then the adjacency matrix of the diagram is given by $(A_{n{-}1})_{i,j}=\delta_{i,j{-}1}+\delta_{i,j{+}1}$. If we start from vertex $1$, the number of closed paths with length $2n{-}4$ is given by the $(1,1)$ component of the matrix $A_{n{-}1}^{2n{-}4}$. It is straightforward to see that $A_{n{-}1}$ has eigenvectors $v_i=\sqrt{\frac 2 n}(\sin(\frac{\pi i}{n}), \sin(\frac{2\pi i}{n}),\ldots,\sin(\frac{(n{-}1) \pi i}{n}))$, and eigenvalues $\lambda_i=2\cos (\frac{\pi i}{n})$, for $i=1,2,\ldots,n{-}1$, thus
\be
C_{n{-}2}=\left(A_{n{-}1}^{2n{-}4}\right)_{1,1}=\sum_{i=1}^{n{-}1}(v_i)_1 \lambda^{2n{-}4}_i (v_i)_1=\frac{2^{2n{-}3}}{n} \sum_{i=1}^{n{-}1}\cos^{2n{-}4}(\frac{i \pi}{n})\sin^2(\frac{i \pi}{n}),
\ee
where note the summand is symmetric under $i\leftrightarrow n{-}i$ (for $n$ odd, the term with $i=(n{-}1)/2$ vanishes), thus one can replace the range of summation by $i=1,\ldots,\lfloor(n{-}1)/2\rfloor$ and multiply the result by $2$. A simple rewriting gives \eqref{Catalan}, which concludes the proof.

One of the most interesting future directions is to consider the scattering equations for general kinematics in light of the special kinematic calculation. It is possible that by rewriting the scattering equations in terms of cross-ratios, one may encounter more general Y-systems. It would be fascinating to find a physical interpretation of such system of equations, especially an interpretation of our formula as certain physical quantities, in analog of the minimal area as the free energy of the Y-system. Doing so may allow us to find simple expressions for the integrand (in particular a proof for the conjecture above), and possibly to connect scattering equations to some integrable system.

\subsection{Comments on the Special Kinematics}

As the first step towards more general kinematics, we study the cases where all kinematic invariants of the form $s_{i,i+m}$ with any fixed $m$ are equal, as long as the diagrams contributing to the double-partial amplitude do not diverge, which is the most general kinematics cyclically symmetric with respect to the canonical ordering. Up to a rescaling, we can always set $s_{i,i+1}=1$ for all $i$, then the simplest cases of this type is $s_{i,i+1}=1$ and $s_{i,i+m}=-1$ for some fixed $m$, while all the remaining $s_{a,b}$ are identically zero (when $n$ is even and $m=(n-2)/2$, we need to set $s_{i,i+m}=-2$ due to momentum conservation). The above case correspond to $m=2$.

However, since these configurations are in general highly singular with respect to the scattering equations, one may encounter problems if evaluating the formula directly on them. For example, in the case $s_{i,i+1}=1$, $s_{i,i+2}=-\frac{1}{3}$ and $s_{i,i+3}=-\frac{2}{3}$ at $7$ points, directly solving the equations shows that apart from three isolated solutions, there is a continuous region of infinite number of solutions, and naively our previous way of evaluating the formula fails. Even for our favorite special kinematic points $m=2$, which has only isolated solutions and they give the correct result, the number of solutions seems to contradict the general counting, $(n-3)!$. To understand these situations properly, one should regard these kinematics as a limit, i.e.~starting by adding to the kinematics small deviations controlled by a scale $\epsilon$ and evaluate the formula, and then take the $\epsilon\rightarrow0$ limit. To illustrate this, here we provide two examples.

First we give an interpretation to the known result for the kinematics $s_{i,i+1}=-s_{i,i+2}=1$ at $6$ points from this point of view. There we get two solutions, and if we fix $\SL2C$ by setting $\{\sigma_1,\sigma_2,\sigma_3\}=\{0,1,-1\}$ they are
\be
\{\sigma_1,\sigma_2,\sigma_3,\sigma_4,\sigma_5,\sigma_6\}=\{0,1,-1,-\frac{1}{2},-\frac{1}{3},-\frac{1}{5}\},\quad
\{0,1,-1,0,1,-1\},
\ee
and the formula gives $-\frac{27}{2}$ on the first solution and $-\frac{1}{2}$ on the second, thus adding up to $-14$ which matches the result calculated from Feynman diagrams. Alternatively, we can start from the configuration
\be
s_{ab}=\left(
\begin{array}{cccccc}
0&1&x&-2-x-y&y&1\\
1&0&1&y&-2-y-z&z\\
x&1&0&1&z&-2-x-z\\
-2-x-y&y&1&0&1&x\\
y&-2-y-z&z&1&0&1\\
1&z&-2-x-z&x&1&0\end{array}
\right),
\ee
where the parameters $x,y,z$ all deviate from $-1$ by some small values proportional to a scale $\epsilon$. This configuration is generic enough to produce all $(n-3)!=6$ solutions, and whatever $\epsilon$ we choose the summation of the formula evaluating on these solutions gives the correct value matching the diagrams. As $\epsilon$ gradually approaches zero, we observe that the evaluation of the formula also approaches zero on three of the solutions. In these solutions, some of the $\sigma_i-\sigma_{i+3}$ become $\mathcal{O}(\epsilon)$, and their role is to compensate the infinitesimal $s_{i,i+3}$ to produce a finite value that keeps the scattering equations satisfied. However, at $\epsilon=0$ terms with $s_{i,i+3}$ disappear, leaving the equations un-balanced, and so these three sets of $\sigma$ values are excluded. For those remaining three solutions, for infinitesimal $\epsilon$ the evaluation of the formula approaches $-\frac{27}{2}$ on one of them and $-\frac{1}{4}$ on the other two, and the two solutions that gives the same value approaches each other. Although collision of $\sigma_i$ and $\sigma_{i+3}$ may still occur in this case, the scattering equations are still balanced at $\epsilon=0$, so all the three solutions remain. Since the multiplicity of the two identified solutions has already been taken into account by the Jacobian of the delta constraints, upon the limit $\epsilon=0$ we only see two distinct solutions and we only need to evaluate the formula once on each to produce the correct result.

For the second example, we look at a more interesting case which gives rise to a continuous region of solutions. This happens first time at $7$ points, when we consider a generic cyclic symmetric configuration, the most general form of which is (after normalizing $s_{i,i+1}=1$)
\be
s_{i,i+1}=1,\quad s_{i,i+2}=-t,\quad s_{i,i+3}=-1+t,\quad \forall i,
\ee
parameterized by a single variable $t$. In this specific example we set $t=\frac{1}{3}$, and we are going to show that from the view of limit the formula still gives correct answer, which is $\frac{462}{25}$. To approach this kinematics from a generic data, we start with the configuration
\be
\begin{split}
&s_{i,i+1}=1,\quad\forall i,\\
&s_{14}=p,\quad s_{25}=q,\quad s_{36}=u,\quad s_{47}=v,\quad s_{15}=x,\quad s_{26}=y,\quad s_{37}=z,\\
&s_{13}=-1-p+q-v+y-z,
\end{split}
\ee
and obtain all the remaining kinematic invariants by momentum conservation, and we set all the seven parameters $p,q,u,v,x,y,z$ to deviate from $-\frac{2}{3}$ by some small values proportional to the scale $\epsilon$. This in general gives all the $(n-3)!=24$ solutions. Again we let $\epsilon$ to approach zero, and we will observe three types of behaviors for the formula: (a) the formula approaches zero on $7$ solutions. The collision of $\sigma$'s again occurs in all these solutions, and at $\epsilon=0$ these solutions are excluded also due to the fact that the equations are no longer balanced. (b) there are $3$ solutions upon which the evaluation of the formula remains finite and the solutions are regular. However, the summation of the formula evaluated just on these solutions does not yet add up to the correct result. (c) the evaluation of the formula diverges on all the remaining $14$ solutions, and these solutions are also regular. However, they always add up to be a finite number and when combined with the evaluation upon the $3$ solutions in case (b) the total summation gives the correct result $\frac{462}{25}$ in the $\epsilon\rightarrow0$ limit!

On the other hand, if we work at $\epsilon=0$ right at the start, we will find only $3$ isolated solutions, which match exactly with the $\epsilon\rightarrow0$ limit of the $3$ solutions in case (b). The $7$ solutions in case (a) are already excluded. Moreover, here we will see a $1$-dimensional continuous region of solutions to the scattering equations, and if we pick up the $14$ solutions in case (c) and take the $\epsilon\rightarrow0$ limit, we will see that they ultimately sit within this region. In other words, upon $\epsilon=0$, the original $14$ isolated solutions emerge into a continuous region of solutions, which also gives a non-trivial contribution to the final result.

This continues to be true for generic $t$, where there is always a continuous region of solutions together with $3$ isolated ones at $\epsilon=0$. Given a certain $\SL2C$ fixing the $3$ isolated solutions in case (b) are actually independent of the value of $t$, and it is possible to work out that the formula evaluated over these $3$ solutions add up to
\be\label{evaluationisolatedsolns}
\frac{14(1-31t+280t^2-543t^3-513t^4+1926t^5-427t^6-1553t^7+863t^8)}{(1-t-2t^2+t^3)^2(1-15t+12t+t^3)^2}.
\ee
On the other hand, the total value determined from the scalar diagrams is
\be
-\frac{14(-4+t)}{(-2+t)^2},
\ee
which only contains one physical pole corresponding to $s_{i,i+1,i+2}=0$, and so all the poles in \eqref{evaluationisolatedsolns} are spurious. Subtracting the above two expressions, we know that the contribution from the continuous region should be
\be\label{evaluationcontinuoussolns}
\frac{14(-1+t)t(1+t)^2(-1+12t-36t^2+27t^3+31t^4-87t^5+113t^6-74t^7+15t^8+t^9)}{(-2+t)^2(1-t-2t^2+t^3)^2(1-15t+12t+t^3)^2}.
\ee
In particular, this contribution vanishes when $t$ assumes one of the roots of its numerator. Actually upon such values one can verify that the continuous region of solutions is absent and all the solutions left are the $3$ isolated solutions, which by themselves determines the correct result. Among these, $t=1$ is what we have mainly studied in this section. If we choose to approach this limit from generic kinematics, we will find that in the neighborhood of the limit the formula remains finit on the $3$ solutions in case (b), while it approaches zero on all the remaining solutions, which are ultimately excluded at $\epsilon=0$.

The above explains that although the formula naively fails for such singular kinematics, it gives the correct answer when regarded as a limit from generic kinematic data. Due to the existence of the continuous region of solutions and its non-trivial contribution to the final result as in \eqref{evaluationcontinuoussolns}, it is interesting to look for a way to determine the contribution from such region directly. A possible solution may be similar to the method proposed in \cite{Beasley:2003fx}.

\section{Consistency Checks}

In this section we perform the two standard consistency checks on tree-level amplitudes. The first is the behavior when one of the particles becomes soft. The second is the presence of a simple pole whenever a kinematic invariant of the form $(\sum_{i\in I} k_i)^2$ for a set $I$ vanishes with residue equaling to the product of lower-point amplitudes.

\subsection{Soft-Limit}

Following the analysis done for Yang-Mills theory in \cite{Cachazo:2013hca} one has that in the limit $k_n^\mu\to \varepsilon {\hat k}_n^\mu$
\ba
{\cal M}^{(0)}_n&\to& \sum_{I=1}^{(n-4)!}\oint_{\Gamma} d\sigma_n\frac{1}{\sum_{a\neq n}\frac{s_{n,a}}{\sigma_{n,a}}}\sum_{\alpha,\beta\in S_{n{-}2}} c_\alpha \tilde c_\beta \frac{\sigma_{\alpha(n{-}1),1}\sigma_{\beta(n{-}1),1}}{\sigma_{\alpha(n{-}1),n}\sigma_{\beta(n{-}1),n}\sigma_{n,1}\sigma_{n,1}}{\cal I}_{n-1}(\alpha|\beta)\nl
&=&\sum_{I=1}^{(n-4)!}\oint_{\Gamma} d\sigma_n\frac{1}{\sum_{a\neq n}\frac{s_{n,a}}{\sigma_{n,a}}}\sum_{i,j=2}^{n{-}1}\frac{\sigma_{i,1}\sigma_{j,1}}{\sigma_{i,n}\sigma_{j,n}\sigma^2_{n,1}}\sum_{\alpha^i,\beta^j\in S_{n{-}3}} c_{\alpha^i} \tilde c_{\beta^j} {\cal I}_{n-1}(\alpha^i,i;\beta^j,j)
\ea
where in addition to the sum over $(n{-}4)!$ solutions for $\sigma_1,\ldots,\sigma_{n{-}1}$ (we have abbreviated their solution-label $I$), the contour $\Gamma$ encircles the $n-3$ zeroes of the first factor in the denominator, and ${\cal I}(\alpha|\beta)$ denote the integrand for $(n{-}1)$-point double-partial amplitude with permutations $\alpha,\beta$; in the second equality we have decomposed the sum over $\alpha$ into the sum over $i\equiv \alpha(n{-}1)$ and the sum over $\alpha^i$, which are permutations within $\{2,\ldots,n{-}1\}\backslash\{i\}$, and similarly for the sum over $\beta$.

Since $\sigma_a$'s are taken to be complex numbers in this paper, the delta functions imposing the scattering equations are in fact poles and all our integrals are contour integrals. In using the residue theorem, one finds that there is no contribution at infinity; for each term in the sum over $i,j$, there is a pole at $\sigma_{n}=\sigma_{1}$, and only for each of the terms with $i=j$, we have a pole at $\sigma_{n}=\sigma_i$ which gives non-zero residue. Upon each pole only one term from $\sum_{a\neq n}\frac{s_{n,a}}{\sigma_{n,a}}$ will contribute to the residue, giving rise to
\be
{\cal M}^{(0)}_n\to\sum_{\textsf{cd}}\left(\frac{1}{s_{n,1}}\sum_{i,j=2}^{n{-}1} f_{\textsf{c}\textsf{a}_i \textsf{a}_n}\tilde f_{\textsf{d}\textsf{b}_j \textsf{b}_n} {\cal M}^{(0)}_{n{-}1}(i^{\textsf{c}\textsf{b}_i}, j^{\textsf{a}_j\textsf{d}}) + \sum_{i=2}^{n{-}1}\frac{1}{s_{n,i}} f_{\textsf{c}\textsf{a}_i \textsf{a}_n}\tilde f_{\textsf{d}\textsf{b}_i \textsf{b}_n} {\cal M}^{(0)}_{n{-}1} (i^{\textsf{cd}})\right),\label{softcolor1}
\ee
where we have pulled out one structure constant of $U(N)$ from ${\bf c}_{\alpha^i}$ and one of $\tilde U(N)$ from $\tilde {\bf c}_{\beta^j}$, and rewritten the remaining factors as full amplitudes ${\cal M}^{(0)}_{n-1}$ with the color indices of particle $i$ (and those of particle $j$ for the first sum inside the bracket) being summed over; note that we have suppressed color indices of other particles, which are $\textsf{a}_l, \textsf{b}_l$ for particle $l$. Let us consider the combination
\ba
&&\sum_{i=1}^{n{-}1}\sum_c  f_{\textsf{c}\textsf{a}_i \textsf{a}_n}{\cal M}^{(0)}_{n{-}1}(1^{\textsf{a}_1\textsf{b}_1}, \ldots,
i^{\textsf{c}\textsf{b}_i},\ldots, (n{-}1)^{\textsf{a}_{n{-}1}\textsf{b}_{n{-}1}}
)\nl=&&\sum_{i=1}^{n{-}1}\sum_c {\rm Tr}([T^{\textsf{a}_i},T^{\textsf{a}_n}] T^{\textsf{c}})\sum_{\alpha\in S_{n{-}2}}{\rm Tr}(T^{\textsf{c}} T^{\textsf{a}_{\alpha(i{+}1)}}\ldots T^{\textsf{a}_{\alpha(i{-}1)}}) M^{(0)}_{n{-}1}(i,\alpha(i{+}1),\ldots,\alpha(i{-}1)),\nl
\ea
where we have used the color-decomposition of ${\cal M}^{(0)}_{n{-}1}$ in the trace basis with the position of particle $i$ fixed. Note that $
\sum_c {\rm Tr}([T^{\textsf{a}_i},T^{\textsf{a}_n}]T^{\textsf{c}}){\rm Tr}(T^{\textsf{c}} \ldots)=-{\rm Tr}(T^{\textsf{a}_n}T^{\textsf{a}_i} \ldots)+{\rm Tr}(T^{\textsf{a}_n} \ldots T^{\textsf{a}_i})$, and by combining the sum over $i$ and that over $\alpha\in S_{n{-}2}$ we get a sum over $\alpha'\in S_{n{-}1}$, \emph{i.e.} permutations of $1,\ldots,n{-}1$; for each permutation there are two terms that differ only by a sign, thus the combination vanishes. This means the sum over $i=2,\ldots,n{-}1$ in the double sum of \eqref{softcolor1} gives minus the term with $i=1$, and similarly for the sum over $j=2,\ldots,n{-}1$, thus \eqref{softcolor1} can be simplified as
\be {\cal M}^{(0)}_n\to \sum_{\textsf{cd}}\sum_{i=1}^{n{-}1}f_{\textsf{c}\textsf{a}_i \textsf{a}_n}\tilde f_{\textsf{d}\textsf{b}_i \textsf{b}_n} \frac{1}{s_{n,i}}  {\cal M}^{(0)}_{n-1} (i^{\textsf{cd}}),\label{softcolor}\ee
which is the correct soft behavior as one can see from Feynman diagrams. This can be compared with the soft limit of Yang-Mills full amplitude,
\be
 {\cal M}^{(1)}_n\to \sum_{\textsf{c}}\sum_{i=1}^{n{-}1} f_{\textsf{c}\textsf{a}_i \textsf{a}_n} \frac{\epsilon_n\cdot k_i}{k_n\cdot k_i}{\cal M}^{(1)}_{n{-}1} (i^{\textsf{c}}).
\ee

The soft limit of the partial amplitude, $M^{(0)}_n(1,\ldots,n)$, can be derived similarly, which can be compared with the soft limit of color-ordered Yang-Mills partial amplitude $M^{(1)}_n(1,\ldots,n)$
\ba &M^{(0)}_n(1,\ldots,n)\to \sum_{\textsf{c}}\left(\frac {f_{\textsf{c}\textsf{a}_1 \textsf{a}_n}} {s_{n,1}}  M^{(0)}_{n{-}1} (1^{\textsf{c}},\ldots,n{-}1)+\frac {f_{\textsf{c}\textsf{a}_{n{-}1} \textsf{a}_n}} {s_{n,n-1}} M^{(0)}_{n{-}1} (1,\ldots,(n{-}1)^{\textsf{c}})\right),\nl
&M^{(1)}_n(1,\ldots,n)\to \left(\frac{\epsilon_n\cdot k_1}{k_n\cdot k_1}+\frac{\epsilon_n\cdot k_{n{-}1}}{k_n\cdot k_{n{-}1}} \right) M^{(1)}_{n{-}1}(1,\ldots,n{-}1).
\ea

\subsection{Factorization}\label{factorization}

For the purpose of showing factorizations of the formula \eqref{uni} in the scalar theory, it sufficies to have a look at only the kinematics singularity defined by
\be\label{factorizationlimit}
k^2_{I_R}=(k_1+k_2+\cdots+k_{n_L})^2\longrightarrow0,
\ee
with $2\leq n_L\leq n-2$, and we denote $L=\{1,\ldots,n_L\}$ and $R$ as its complement set, with $n_R=n-n_L$. Upon such a limit, only a subset of $(n_L-2)!\times(n_R-2)!$ from all the solutions are singular \cite{Cachazo:2013gna}. To make factorization manifest, we start by choosing a special redefinition
\be
\sigma_a=\frac{s}{u_a},\quad a\in L,\quad\quad\quad
\sigma_a=\frac{v_a}{s},\quad a\in R,
\ee
where we regard $v_{n-1}$ as being fixed to a specific value $v^*_{n-1}$, and leave $s$ as well as the remaining $u$'s and $v$'s as variables to be integrated over. Now the measure transforms to
\be\label{changeofmeasure1}
\prod_{a=1}^{n}d\sigma_a=(-1)^{n_L+1}s^{n_L-n_R-1}\frac{v_{n-1}}{(\prod u)^2}ds\prod_{a\in L}du_a\prod_{a\in R\backslash\{n-1\}}dv_a.
\ee
where $\prod u$ denotes the product of all $u_a$ with $a\in L$. We choose to fix $\{u_1,u_2,v_n\}$ to get rid of the $\SL2C$ redundancy. Faddeev-Popov method in this gauge-fixing gives rise to a Jacobian~\footnote{For explicit derivations via Faddeev-Popov gauge-fixing, please refer to the complementary notes on \url{http://ellisyeyuan.wordpress.com/2013/07/07/soft-limits-and-factorizations/}.}
\be\label{faddeevpopov}
-\frac{2u_{1,2}v_{n-1,n}(-s^4+u_1u_2v_{n-1}v_{n})}{s^2v_{n-1}}.
\ee
If $s$ becomes infinitesimal in the neighborhood of singular solutions, we are able to approximate \eqref{faddeevpopov} and we can think of the $u_1$ in the parentheses above as $(u_1-0)$ where the ``$0$'' corresponds to the puncture of the internal particle in the factorization (the same for $u_2,v_{n-1},v_n$).

The reason that $s$ is constrained to be infinitesimal near singular solutions can be justified by the behavior of the delta constraints. Here we choose to eliminate the constraints corresponding to particles $\{1,2,n\}$. Then for $a\in R$, we can expand the $n_R-1$ constraints with respect to $s$
\be\label{rightconstraints}
s\left(\frac{s_{a,I_R}}{v_a}+\frac{s_{a,n_L+1}}{v_{a,n_L+1}}+\cdots+\frac{s_{a,n}}{v_{a,n}}\right)+\mathcal{O}(s^3)=0.
\ee
However, we only need $n_R-2$ of them for the right sub-amplitude, and so one delta function needs to be isolated to produce a constraint on the invariant mass $k^2_{I_R}$, so that when it approaches zero a massless internal particle is emerged. For this purpose, we dress each constraint in \eqref{rightconstraints} with a factor $\frac{v_av_{n,a}}{sv_n}$ and sum them up, we obtain
\be\label{kiconstraint}
-k^2_{I_R}+s^2\left(F(k,u,v)+\mathcal{O}(s)\right)=0,
\ee
with $F$ some function independent of $s$. From this, we explicitly see that upon the factorization limit \eqref{factorizationlimit} we do have a subset of solutions which gives $s\rightarrow0$. In producing the constraint \eqref{kiconstraint} from the original ones we get an additional Jacobian $\frac{v_{n-1}v_{n,n-1}}{sv_{n}}$. Moreover, for $a\in L$, the $n_L-2$ delta constraints are expanded to
\be\label{constraintsright}
-\frac{u^2_a}{s}\left(\frac{s_{a,1}}{u_{a,1}}+\cdots+\frac{s_{a,n_L}}{u_{a,n_L}}+\frac{s_{a,I_L}}{u_a}\right)+\mathcal{O}(s)=0.
\ee

Given \eqref{kiconstraint}, now we focus only in the regions where $s$ is constrained to be infinitesimal, we are allowed to pick up only the leading terms in \eqref{faddeevpopov}, \eqref{rightconstraints}, \eqref{kiconstraint} and \eqref{constraintsright}. Collecting these results together with \eqref{changeofmeasure1}, we conclude that the measure and the delta constraints approximates to
\be\label{measureconstraints}
\begin{split}
-ds^2\prod_{a=3}^{n_L}du_a\prod_{a=n_L+1}^{n-2}dv_a
\frac{(u_{1,2}u_1u_2v_{n-1,n}v_{n-1}v_n)^2}{(\prod u)^4}s^{2n_L-2n_R-6}\delta(s^2F-k^2_{I_R})\cdot\\
\cdot{\prod_{a\in L\cup\{I_L\}\backslash\{1,2\}}}\delta(\sum_{b\in L\cup\{I_L\}\backslash\{a\}}\frac{s_{a,b}}{u_{a,b}})
{\prod_{a\in R\cup\{I_R\}\backslash\{n-1,n\}}}\delta(\sum_{b\in R\cup\{I_R\}\backslash\{a\}}\frac{s_{a,b}}{v_{a,b}}).
\end{split}
\ee

Then we go on with the summation of color-dressed Parke-Taylor factors. It is convenient to start with the form as in \eqref{chaindecomposition1} and \eqref{chaindecomposition2} where the color factors are expressed in terms of structure contants.
%in which we fix one particle in $L$ and one particle in $R$ at the ends of the color chain , say $1$ and $n$, and all the remaining particles fully permuted
%\be
%\sum_{\gamma\in S_{n-2}}\frac{f_{\textsf{a}_1,\textsf{a}_{\gamma(2)},\textsf{b}_1}f_{\textsf{b}_1,\textsf{a}_{\gamma(3)},\textsf{b}_2}\cdots f_{\textsf{b}_{n-3},\textsf{a}_{\gamma(n-1)},\textsf{a}_n}}{\sigma_{1,\gamma(2)}\sigma_{\gamma(2),\gamma(3)}\cdots\sigma_{\gamma(n-1),n}\sigma_{n,1}}.
%\ee
With infinitesimal $s$, it is not hard to check that the leading terms are contributed solely by terms where both the two Parke-Taylor forms are of the pattern such that all labels in $L$ sit in front of all the labels in $R$, which behave as
\be\label{parketaylorbehavior}
\begin{split}
&\frac{1}{\sigma_{1,\alpha(2)}\cdots\sigma_{\alpha(n_L),\beta(n_L+1)}\cdots\sigma_{\beta(n-1),n}\sigma_{n,1}}\longrightarrow\\
&\quad\quad\frac{(-1)^{n_L}s^{-n_L+n_R+2}(\prod u)^2}{(u_1u_{1,\alpha(2)}\cdots u_{\alpha(n_L-1),\alpha(n_L)}u_{\alpha(n_L)})(v_{\beta(n_L+1)}v_{\beta(n_L+1),\beta(n_L+2)}\cdots v_{\beta(n-1),n}v_n)},
\end{split}
\ee
where $\alpha$ is any permutation within the label set $L\backslash\{1\}$, and $\beta$ any permutation within the label set $R\backslash\{n\}$. All the other Parke-Taylor forms are of higher order in $s$ compared to \eqref{parketaylorbehavior}, and so they are irrelavent in the factorization. For these leading terms, it is clear that each copy of their corresponding color factors also breaks into two parts, with one new fixed end arising in each part, and the two new ends are glued by a Kronecker delta with indices in the adjoint representation of the color groups
\be
\begin{split}
&\sum_{\{\textsf{c}\}}f_{\textsf{a}_1,\textsf{a}_{\alpha(2)},\textsf{c}_1}\cdots f_{\textsf{c}_{n_L-2},\textsf{a}_{\alpha(n_L)},\textsf{c}_{n_L-1}}f_{\textsf{c}_{n_L-1},\textsf{a}_{\beta(n_L+1)},\textsf{c}_{n_L}}\cdots f_{\textsf{c}_{n-3},\textsf{a}_{\beta(n-1)},\textsf{a}_n}=\\
&\quad\sum_{\textsf{a}_L,\textsf{a}_R}(\sum_{\{{\textsf{c}'}\}}f_{\textsf{a}_1,\textsf{a}_{\alpha(2)},{\textsf{c}'}_1}\cdots f_{{\textsf{c}'}_{n_L-2},\textsf{a}_{\alpha(n_L)},\textsf{a}_L})\delta_{\textsf{a}_L,\textsf{a}_R}
(\sum_{\{{\textsf{c}''}\}}f_{\textsf{a}_R,\textsf{a}_{\beta(n_L+1)},{\textsf{c}''}_{n_L}}\cdots f_{{\textsf{c}''}_{n-3},\textsf{a}_{\beta(n-1)},\textsf{a}_n}).
\end{split}
\ee
As a consequence, each copy of the summation over Parke-Taylor forms exactly splits into the product of summations on the left part and that on the right part. Combining two copies of this summation together with \eqref{measureconstraints}, it is easy to see that if we regard the factors $\{u_1,u_2,v_{n-1},v_n\}$ therein as $\{u_1-u_{I_L},u_2-u_{I_L},v_{n-1}-v_{n_R},v_n-v_{n_R}\}$, with $u_{I_L}$ and $v_{I_R}$ as punctures for the internal particles on the left part and right part, which are fixed to be zero, then we have an emergent $\SL2C\times\SL2C$ redundancy in these leading terms, acting on $L\cup\{u_{I_L}\}$ and $R\cup\{v_{I_R}\}$ respectively. When we integrate $s^2$ out, we see that $\mathcal{M}^{(0)}_n$ factorizes in the correct way
\be
\mathcal{M}^{(0)}_n\longrightarrow \sum_{\textsf{a}_L,\textsf{a}_R,\textsf{b}_L,\textsf{b}_R}\mathcal{M}^{(0)}_{n_L+1}(1,\ldots,n_L,I_L^{\textsf{a}_L,\textsf{b}_L})\frac{-\delta_{\textsf{a}_L,\textsf{a}_R}\delta_{\textsf{b}_L,\textsf{b}_R}}{k_I^2}\mathcal{M}^{(0)}_{n_R+1}(I_R^{\textsf{a}_R,\textsf{b}_R},n_L+1,\ldots,n).
\ee

We can also apply the above discussion to the double-partial amplitudes $m^{(0)}(\alpha|\beta)$ with any orderings $\alpha$ and $\beta$. By \eqref{parketaylorbehavior}, we see that $m^{(0)}(\alpha|\beta)$ will have diverging leading terms in the form
\be\label{mfactorization}
m^{(0)}_n(\alpha|\beta)\longrightarrow m^{(0)}_{n_L+1}(\alpha_L,I_L|\beta_L,I_L)\frac{-1}{k^2_I}m^{(0)}_{n_R+1}(I_R,\alpha_R|I_R,\beta_R)
\ee
if and only if the labels contained in $k_I$ forms a consecutive subset in both $\alpha$ and $\beta$, and since $m^{(0)}(\alpha|\beta)$ is a function of only kinematic invariants, it vanishes whenever there doesn't exist such a factorization channel. By studying factorizations recursively it is straightforward to observe that up to a sign $m^{(0)}(\alpha|\beta)$ has the form as presented in \eqref{sumscalardiagrams}. For the overall sign, we can just pick up any tree diagram $g\in m^{(0)}(\alpha|\beta)$ and do $n-3$ consecutive factorizations according to the propagators therein to fully factorize $m^{(0)}(\alpha|\beta)$ down to $n-2$ cubic vertices. On the one hand \eqref{mfactorization} indicates that each propagator gives rise to a minus sign, and on the other hand as in \eqref{simpleexamplesD} it is easy to see that each cubic vertex gives $+1$ if the $\alpha$ and $\beta$ orderings of the three labels are the same and $-1$ if flipped, so the overall sign is $(-1)^{n-3+n_{\text{flip}}}$. Since as argued in Section 3, this value is independent of the diagram $g$ we choose, we have thus verified \eqref{sumscalardiagrams}.

\section{Discussions}

In this paper we provided a unified description of the tree-level S-matrix of a colored massless cubic scalar theory, Yang-Mills and gravity. The new description manifests a relation between factors that contain the color information and factors that contain the polarization information. All amplitudes are written as integrals over the moduli space of an $n$-punctured sphere. The locations of the punctures are fixed by solving the scattering equations which in general give rise to $(n-3)!$ solutions. Amplitudes are then obtained as a sum over solutions of an integrand and a Jacobian factor
\be
{\cal M}^{({\bf s})}_n = \sum_{I=1}^{(n-3)!}\left.\frac{{\cal I}^{({\bf s})}}{{\rm det}'\Phi}\right|_I
\ee
where ${\cal I}^{({\bf s})}$ with ${\bf s} =0,1,2$ represent the integrand for scalars, gluons or gravitons respectively. The integrands, in their simplest forms, are given by
\be
{\cal I}^{({ 0})} = C_{U(N)}^2 , \quad {\cal I}^{({1})} = C_{U(N)} E_{\epsilon}, \quad {\cal I}^{({2})} = E_{\epsilon}^2.
\ee
This shows that solution by solution one has $({\cal I}^{({ 1})})^2 = {\cal I}^{({ 2})} {\cal I}^{({ 0})}$. This connection between the square of Yang-Mills and the product of gravity with a $\phi^3$ theory has the same structure as that found by Hodges using twistor diagrams in \cite{Hodges:2011wm}.

Very nicely, there is a third way to make this connection between Yang-Mills, gravity and $\phi^3$ explicit. Recall that the KLT construction represents a gravity amplitude as a linear combination of products of partial Yang-Mills amplitudes. The sum is over certain pairs of $(n-3)!$ permutations and it is given schematically as
\be
{\cal M}^{({ 2})}_n = \sum_{\alpha , \beta\in S_{n-3}} M_n^{(1)}(\alpha )S(\alpha|\beta )M_n^{(1)}(\beta ).
\ee
In section 3.1 we proved that if $S(\alpha |\beta )$ is taken to be the entries of an $(n-3)!\times (n-3)!$ matrix, $S_{\rm KLT}$, then
\be
S_{\rm KLT} = (m_{\rm scalar})^{-1}
\label{hum}\ee
where the entries of $m_{\rm scalar}$ are given by the double partial amplitudes of the scalar theory, $m(\alpha|\beta)$. Therefore one has
\be
{\cal M}^{(2)}_n = \sum_{\alpha , \beta} M_n^{(1)}(\alpha )(m_{\rm scalar}^{-1})^{\alpha}_{\beta}M_n^{(1)}(\beta ).
\ee
It would be interesting to explore possible connections among all three descriptions.

In the work of Broedel, Schlotterer and Stieberger \cite{Broedel:2013tta}, it was shown that the field theory limit of certain string theory integrals on the disk compute the entries of $S_{\rm KLT}^{-1}$. Combining this with our result \eqref{hum} one can conclude that
\be
m^{(0)}_n(\gamma |\beta) =\!\!\! \left.\int_\gamma [d^nz]\frac{ \prod_{i<j}|z_i-z_j|^{\alpha' s_{ij}}}{(z_{1}-z_{\beta(2)})(z_{\beta(2)}-z_{\beta(3)})\cdots (z_{\beta(n-2)}-z_{n})(z_n-z_{n-1})(z_{n-1}-z_1)}\right|_{\alpha'\to 0}
\nonumber\ee
where the dependence on the permutation $\gamma$ is through the region of integration given by
\be
z_1<z_{\gamma(2)}<\ldots < z_{\gamma(n-2)}<z_{n-1}<z_n
\ee
and the measure $[d^nz]$ is defined to be $d^nz/{\rm vol}(SL(2,\mathbb{R}))$.

Recall the formula for $m^{(0)}_n(\alpha |\beta)$ found in this work is given by
\be
\!\!\int\!\! \frac{d\,^n\sigma }{\textrm{vol}\,\SL2C} \frac{\prod_a {}'\delta(\sum_{b\neq a} \frac{s_{ab}}{\sigma_{a b}})}{(1,\alpha(2),\alpha(3),\ldots, \alpha(n-2),n-1,n)(1,\beta(2),\beta(3),\ldots, \beta(n-2),n,n-1)}
\nonumber\ee
with $(a_1,a_2,\ldots, a_n) =(\sigma_{a_1}-\sigma_{a_2})\cdots (\sigma_{a_n}-\sigma_{a_1})$.

Finding a direct proof of the equivalence of the two formulas for $m^{(0)}_n (\alpha |\beta)$ is an important problem which might also give a reason why the formulas for Yang-Mills and gravity amplitudes are strikingly similar to those of string amplitudes in the Gross-Mende or high energy limit \cite{Gross:1987ar}. Hints in this direction were already discussed in \cite{Cachazo:2013gna}.

Clearly there are other pressing issues which we leave for future work. The first is finding a formula analogous to the reduced Pfaffian, ${\rm Pf}'\Psi$, which could accommodate the scattering of fermions. In dimensions less than twelve for gravity and less than eleven for Yang-Mills one could try and find a supersymmetric version of the reduced Pfaffian. In four dimensions, our formula has been shown to agree with the Witten-RSV formula for the scattering of gluons for up to eight particles and in all helicity sectors \cite{Cachazo:2013hca}. The agreement happens solution by solution of the scattering equations which means that the formulas are equivalent at the level of the integrand. It is well known that the Witten-RSV formula for gluons is only a set of components of an elegant ${\cal N}=4$ supersymmetric formula~\cite{Witten:2003nn,Roiban:2004yf}. An analogous formulation for ${\cal N}=8$ supergravity is also known~\cite{Cachazo:2012da,Cachazo:2012kg}. This suggests that a supersymmetric generalization of the Pfaffian exists in four dimensions and perhaps in higher. Finally, all the formulas obtained so far have been restricted to tree level S-matrices. Extending this formalism to loop level will likely involve higher genus Riemann surfaces and constrains on the theories. It would be fascinating to understand the class of theories whose S-matrices can be recast in this way.

%\appendix
%\section{}

\begin{acknowledgments}
The authors would like to thank Yu-tin Huang and Oliver Schlotterer for useful discussions and comments on the manuscirpt. This work is supported by Perimeter Institute for Theoretical Physics. Research at Perimeter Institute is supported by the Government of Canada through Industry Canada and by the Province of Ontario through the Ministry of Research \& Innovation.
\end{acknowledgments}

\bibliographystyle{JHEP}
\bibliography{ScatteringEquations}

\end{document}